\documentclass{aa}
\usepackage[varg]{txfonts}
\usepackage{natbib}  
\usepackage{graphicx}
\usepackage{txfonts}
\bibpunct{(}{)}{;}{a}{}{,}
\usepackage{tabularx}
\begin{document} 
   \title{Structure and stability in TMC-1: analysis of NH$_3$ molecular line and \textit{Herschel} continuum data}
\titlerunning{Structure and stability in TMC-1}
\authorrunning{Feh\'er et al.}
   \author{O. Feh\'er\inst{1,2},
   				L. V. T\'oth\inst{1},
				D. Ward-Thompson\inst{3},					
					J. Kirk\inst{3},
					A. Kraus\inst{4},
					V.-M. Pelkonen\inst{5},
					S. Pintér\inst{1},
					S. Zahorecz\inst{1, 6}}

   \institute{
   E\"otv\"os Lor\'and University, Department of Astronomy, P\'azm\'any P\'eter s\'et\'any 1/A, 1117 Budapest, Hungary \and
   Konkoly Observatory, Research Centre for Astronomy and Earth Sciences, Hungarian Academy of Sciences, H-1121 Budapest, Konkoly Thege Mikl\'os \'ut 15-17, Hungary \and
   Jeremiah Horrocks Institute, University of Central Lancashire, Preston PR1 2HE, UK \and
   Max Planck Institute for Radioastronomy, Auf dem Hügel 69, 53121 Bonn, Germany \and
   Department of Physics, PO Box 64, 00014 University of Helsinki, Finland \and
   European Southern Observatory, Karl-Schwarzschild-Str. 2, D-85748, Garching bei M\"unchen, Germany
             }

  \abstract
    {}
   {We examined the velocity, density and temperature structure of Taurus Molecular Cloud-1 (TMC-1), a filamentary cloud in a nearby quiescent star forming area, to understand its morphology and evolution.}
	{We observed high S/N, high velocity resolution NH$_3$(1,1) and (2,2) emission on an extended map. By fitting multiple hyperfine-split line profiles to the NH$_3$(1,1) spectra we derived the velocity distribution of the line components and calculated gas parameters on several positions. \textit{Herschel SPIRE} far-infrared continuum observations were reduced and used to calculate the physical parameters of the \textit{Planck} Galactic Cold Clumps (PGCCs) in the region, including the two in TMC-1. The morphology of TMC-1 was investigated with several types of clustering methods in the position-velocity-column density parameter space.}
	{Our \textit{Herschel}-based column density map shows a main ridge with two local maxima and a separated peak to the south-west. H$_2$-column densities and dust colour temperatures are in the range of 0.5-3.3\,$\times$\,10$^{22}$\,cm$^{-2}$ and 10.5-12\,K, respectively. NH$_3$-column densities and H$_2$-volume densities are in the range of 2.8-14.2\,$\times$\,10$^{14}$\,cm$^{-2}$ and 0.4-2.8\,$\times$\,10$^4$\,cm$^{-3}$. Kinetic temperatures are typically very low with a minimum of 9\,K at the maximum NH$_3$ and H$_2$-column density region. The kinetic temperature maximum was found at the protostar IRAS 04381+2540 with a value of 13.7\,K. The kinetic temperatures vary similarly as the colour temperatures in spite of the fact that densities are lower than the critical density for coupling between the gas and dust phase.
     The k-means clustering method separated four sub-filaments in TMC-1 with masses of 32.5, 19.6, 28.9 and 45.9 M$_{\sun}$ and low turbulent velocity dispersion in the range of 0.13-0.2\,kms$^{-1}$.}
	{The main ridge of TMC-1 is composed of four sub-filaments that are close to gravitational equilibrium. We label them TMC-1F1 through F4. TMC-1F1, TMC-1F2 and TMC-1F4 are very elongated, dense and cold. TMC-1F3 is a little less elongated and somewhat warmer, probably heated by the Class I protostar, IRAS 04381+2540 that is embedded in it. TMC-1F3 is approximately 0.1\,pc behind TMC1-F1. Because of its structure, TMC-1 is a good target to test filament evolution scenarios.}
   {}
   \maketitle

\section{Introduction}

Large-scale filaments have long been recognized as fundamental environments of the star forming process. Several studies revealed that filaments are common structures in interstellar clouds with young stars in different stages of formation located in them \citep{schneider1979, bally1987, hatchell2005, goldsmith2008}. Recent results based on \textit{Herschel} FIR observations of nearby star forming regions directed the attention again to the connection
between filaments, dense cores and stars \citep[see e.g.][]{andre2010}. Since star formation occurs mostly in prominent filaments, characterizing the physical properties of these regions is the key to understand the process of star formation.

The Taurus Molecular Cloud is one of the closest, low-mass star forming regions at 140\,pc \citep{elias1978, onishi2002}. It was a target of several cloud evolution and star formation studies \citep{ungerechts1987, mizuno1995, goldsmith2008} and was mapped extensively in CO \citep{ungerechts1987, onishi1996, narayanan2008} and extinction \citep{cambresy1999, padoan2002, dobashi2005}. The most massive molecular cloud in Taurus is the Heiles Cloud 2 \citep[HCL 2;][]{heiles1968, onishi1996, toth2004} which is the second brightest source on the 353, 545 and 857\,GHz \textit{Planck} maps \citep{tauber2010, planck2011a} in Taurus, with a flux density peak in \object{HCL 2B} \citep{heiles1968}. 

\begin{figure}[!t]
\centering
\includegraphics[width=\linewidth, trim=0 0 0 0cm]{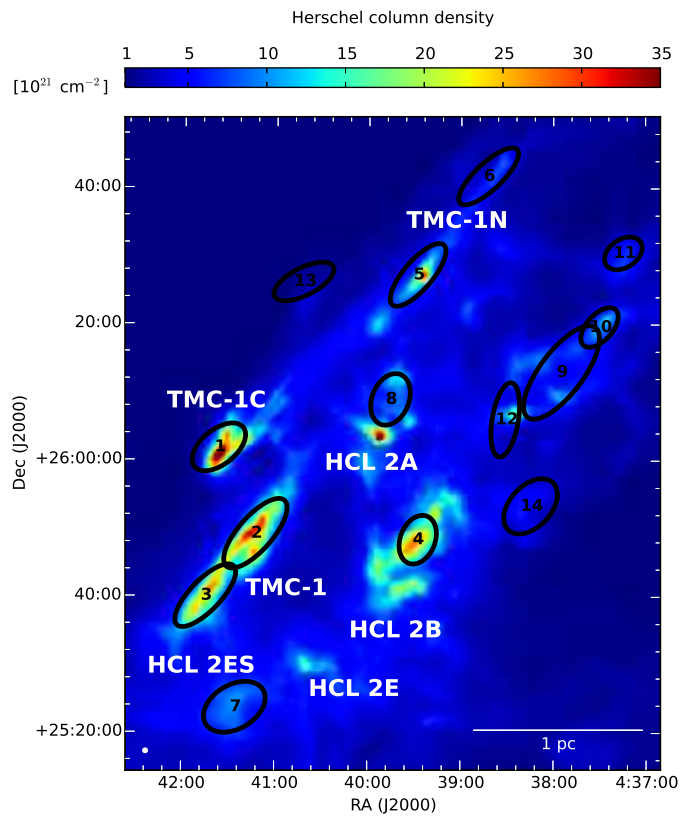}
\caption{\textit{Herschel} N(H$_2$) column density map of HCL 2 and the main parts of the cloud. PGCCs with flux quality of 1 and 2 are plotted with black ellipses and the numbers refer to Table~\ref{pgcc2} that lists the properties of the PGCCs.}
\label{hcl22}
\end{figure}

There are 406 \textit{Planck} Galactic Cold Clumps (PGCCs) in the Taurus region \citep{planck2015}. These are clustered and form 76 groups as it was found by \citet{toth2016} using the Minimum Spanning Tree method \citep{planck2011c}. One of the largest PGCC groups is \object{HCL 2} with 14 cold clumps, their parameters are listed in Table \ref{pgcc2}. The clump temperatures and densities were derived from \textit{Herschel} maps (see Section \ref{herscheldata}). The structure of the PGCC group is shown in Fig. \ref{hcl22} that also serves as a finding chart for the parts of HCL 2.

Molecular emission maps show the south-eastern part of HCL 2 as a ring-like structure called the Taurus Molecular Ring \citep[TMR;][]{schloerb1984}. TMC-1 appears as a dense, narrow ridge at the eastern edge of \object{TMR}. Two PGCCs are found in this region, almost parallel with the galactic plane. Radio spectroscopic surveys indicated a considerable variation in the relative abundance of gas phase molecules, e.g. well-separated cyanopolyyne, ammonia \citep{little1979} and SO \citep{pratap1997} peaks (see also Table \ref{pgcc2}). Studies of NH$_3$, HC$_\mathrm{n}$N (n\,=\,3,5,7), C$_4$H, CS, C$^{34}$S, HCS$^+$ \citep{toelle1981, gaida1984, olano1988, hirahara1992}, CH \citep{suutarinen2011, sakai2012} and other carbon-chain and sulfur-containing molecules \citep{langer1995, pratap1997} show complex velocity and chemical structure in TMC-1. \citet{hirahara1992} suggested that the displacement of the column density peaks of carbon-chain and nitrogen bearing molecules is due to the different evolutionary states of the regions. The molecular abundances in TMC-1 were modelled by \citet{mcelroy2013} using the fifth release of the  UMIST Database for Astrochemistry.

The morphologic and evolutionary connection between TMC-1 and HCL 2 was described by \citet{schloerb1983} and \citet{schloerb1984} as TMC-1 being a collapsing, fragmented part of a torus shaped cloud structure (TMR). \citet{cernicharo1987} considered TMC-1 a clump inside a thick, bent filamentary complex projecting itself as a loop on the plane of the sky. The existence and relative motion of fragments in TMC-1 were discussed in various papers \citep{snell1982, schloerb1983, olano1988, hirahara1992, langer1995, pratap1997}. \citet{nutter2008} concluded that in sub-millimeter and FIR, TMR is composed of a dense filament on one side (TMC-1) and a series of point-like sources on the other. High spatial resolution \textit{Herschel} \citep{pilbratt2010} observations resolved the cold clumps found by \textit{Planck} and their physical characteristics could be determined. \citet{malinen2012} identified two long filaments in HCL 2 based on NIR extinction and \textit{Herschel} data, one of which was TMC-1. They fitted the column density perpendicular to the filament with a Plummer-like profile and derived an average width of 0.1\,pc, which is typical of star forming filaments.

In this paper we discuss the substructures of TMC-1 using our high S/N NH$_3$(1,1) and NH$_3$(2,2) line observations and high resolution \textit{Herschel} FIR maps. We examine the possibility of overlapping sub-filaments being present inside the cloud by partitioning it using various clustering methods in the derived parameter space. We reveal the morphology of TMC-1 and assess the physical state and stability of the different cloud parts.

\section{Observations and data reduction}
\begin{table*}[htpb]
\caption{Properties of the PGCCs in HCL 2. See Section \ref{herschelcalc} for clump temperatures and densities, and Fig. \ref{hcl22} as a finding chart.}
	\centering
		\begin{tabular}{rcccccccl} 
		\hline
 \multicolumn{1}{c}{N} & \multicolumn{1}{c}{PGCC clump} & \multicolumn{1}{c}{$\alpha$(2000)} & \multicolumn{1}{c}{$\delta$(2000)} & \multicolumn{1}{c}{T$\mathrm{_{dust, min}}$} & \multicolumn{1}{c}{T$\mathrm{_{dust, ave}}$} & \multicolumn{1}{c}{N(H$_2$)$\mathrm{_{peak}}$} & \multicolumn{1}{c}{N(H$_2$)$\mathrm{_{ave}}$} & Other IDs \\	
 &  & \multicolumn{1}{c}{[deg]} & \multicolumn{1}{c}{[deg]} & \multicolumn{1}{c}{[K]} & \multicolumn{1}{c}{[K]} & \multicolumn{2}{c}{[10$^{22}$ cm$^{-2}$]} & \\
\hline
\hline
 1 & G174.09-13.24 & 70.4039 & 26.0315 &  9.9 & 12.4 & 4.1 &  0.7 & TMC-1C $\mathrm{^{(1)}}$\\
 2 & G174.20-13.44 & 70.3034 & 25.8207 & 10.4 & 12.7 & 3.3 &  0.7 & TMC-1 $\mathrm{^{(2)}}$, TMC-1(NH3) $\mathrm{^{(3)}}$\\
 3 & G174.40-13.45 & 70.4373 & 25.6680 & 10.8 & 13.0 & 2.8 &  0.7 & TMC-1 $\mathrm{^{(2)}}$, TMC-1 CP $\mathrm{^{(3,4)}}$\\
 4 & G173.95-13.76 & 69.8655 & 25.8070 & 11.2 & 12.5 & 3.0 & 1.1 & HCL 2B $\mathrm{^{(2)}}$\\
 5 & G173.44-13.34 & 69.8673 & 26.4553 & 10.5 & 12.7 & 3.1 &  0.5 & \object{TGU H1211 P6} $\mathrm{^{(5)}}$, \object{TMC-1N} $\mathrm{^{(6)}}$\\
 6 & G173.14-13.32 & 69.6767 & 26.6971 & 11.8 & 12.9 &  0.9 &  0.3 &  TMC-1N $\mathrm{^{(6)}}$\\
 7 & G174.57-13.68 & 70.3566 & 25.3955 & 12.3 & 13.3 & 1.1 &  0.6 & \object{HCL 2ES} $\mathrm{^{(4)}}$\\
 8 & G173.73-13.49 & 69.9406 & 26.1503 & 11.3 & 12.7 & 2.7 &  0.6 & \object{TGU H1211 P5} $\mathrm{^{(5)}}$\\
 9 & G173.40-13.76 & 69.4780 & 26.2173 & 11.5 & 12.7 & 1.2 &  0.4 & \object{TGU H1211 P4} $\mathrm{^{(5)}}$\\
10 & G173.25-13.76 & 69.3752 & 26.3275 & 11.3 & 12.5 & 1.3 &  0.4 & ...\\
11 & G173.07-13.69 & 69.3107 & 26.5089 & 11.8 & 12.5 &  0.8 &  0.4 & \object{[WMD94] Tau A 11} $\mathrm{^{(7)}}$\\
12 & G173.58-13.73 & 69.6301 & 26.1011 & 11.0 & 12.8 & 1.7 &  0.5 & ...\\
13 & G173.64-13.14 & 70.1768 & 26.4379 & 11.3 & 12.9 &  0.5 &  0.2 & ...\\
14 & G173.71-13.91 & 69.5612 & 25.8895 & 12.4 & 13.1 &  0.7 &  0.3 & ...\\
\hline
		\end{tabular}
	 \tablebib{(1) \citet{myers1983}; (2) \citet{heiles1968}; (3) \citet{little1979}; (4) \citet{toth2004}; (5) \citet{dobashi2005}; (6) \citet{malinen2012};  (7) \citet{wood1994}.}
	\label{pgcc2}
\end{table*}

\subsection{NH$_3$ observations}
\label{ammonia}

\begin{figure}[b]
	\centering
		\includegraphics[angle=-90, trim=1cm 1cm 3cm 0cm, width=1.1\linewidth]{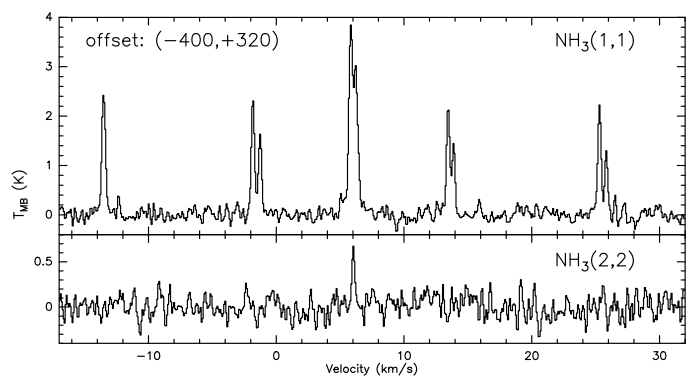}
\caption{Typical NH$_3$(1,1) and NH$_3$(2,2) spectra on our NH$_3$ grid. This position has the highest NH$_3$(1,1) optical depth from the spectra where we observed NH$_3$(2,2) line as well (position \#5 in Table \ref{amm}).}
		\label{line_example}
\end{figure}

The observations of the NH$_3$(J,K)\,=\,(1,1) and (2,2) inversion transitions were carried out with the Effelsberg 100-m telescope\footnote{The 100-m telescope at Effelsberg/Germany is operated by the Max-Planck-Institut für Radioastronomie on behalf of the Max-Planck-Gesellschaft (MPG)} during November 21-30, 2008, April 8-13, 2015 and September 10, 2015. The map was obtained in raster mode, which was a number of targeted observations on a grid with a spacing of 40\,$\arcsec$ (0.027\,pc at the distance of TMC-1) with a velocity resolution of 0.038\,kms$^{-1}$. The (0,0) position of the spectral grid was $\alpha$(2000)\,=\,04:41:42.5, $\delta$(2000)\,=\,25:41:27. The spectra from 2008 were observed in both position switching (PSW) and frequency switching (FSW) modes, both lines and polarizations were measured simultaneously. 44\% of the data were PSW and 56\% FSW, the frequency throw of the FSW observations was 5\,MHz. After the folding and baseline subtraction of the FSW spectra, the FSW and PSW observations were averaged together weighted by noise. The spectra from 2015 were measured only in FSW mode with a frequency throw of 6\,MHz. The pointing errors are within 5-10\,$\arcsec$ for both datasets. We observed a total of 258 positions with a typical integration time of 4\,minutes, a sample spectrum is shown in Fig. \ref{line_example}.

The observing sessions started by measuring the continuum emission of calibration sources (\object{NGC\,7027}, \object{W3OH}). Pointing measurements towards a source close to the target were repeated at least once per hour. For the observations from 2008 NGC\,7027 was used for flux calibration with an assumed flux density of 5.51\,Jy \citep{zijlstra2008}. With the average beam size of 37.1\,$\arcsec$ of the receiver this is equivalent to a main beam brightness temperature T$\mathrm{_{MB}}$\,=\,8.7\,K. We calibrated to the elevation dependent atmospheric and instrumental gain and efficiency with a quadratic function, which was calculated from observing NGC\,7027 at different elevations during the observing sessions. The calibration of the 2008 dataset is estimated to be accurate to $\pm$15\%.

The data from 2015 was observed with a different receiver (the secondary focus system instead of the prime focus K-band receiver used in 2008) which had an average beam size of 36.6\,$\arcsec$. The flux calibrator was NGC\,7027 again, but since is a planetary nebula that expands adiabatically, its flux density decreases. According to the model by \citet{zijlstra2008} it was $\approx$\,5.46\,Jy in 2015. This gives T$\mathrm{_{MB}}$\,=\,8.8\,K which was used to calibrate the 2015 dataset. The spectra on the positions observed in both 2008 and 2015 have an $\approx$10\% difference in intensity, resulting in a 15\% accuracy of the merged dataset. The data were reduced and further analysed with the software packages CLASS and GREG\footnote{http://www.iram.fr/IRAMFR/GILDAS}. The calibrated spectra have a typical rms noise of 0.15\,K. NH$_3$(1,1) was detected with S/N\,>\,3 on 171 positions and the NH$_3$(2,2) line was detected with S/N\,>3\, on 17 positions. The given line temperatures in this paper are in T$\mathrm{_{MB}}$.

\subsection{Separating NH$_3$(1,1) velocity components}
\label{nh3sep}

In order to discern the different velocity components in the NH$_3$(1,1) data, instead of fitting one hyperfine-split (HFS) line profile, we analyzed the spectra as follows. We first fitted a single HFS line profile, subtracted it, then searched the residual for a second component with S/N\,>\,3. This line was also fitted with a HFS line profile. Since in some cases the HFS structure of the secondary components were not detected and the $\tau$(1,1) optical depth could not be determined well, the v$\mathrm{_{LSR}}$ line velocity in the Local Standard Rest frame and the peak T$\mathrm{_{MB}}$ values from this fit were verified with a Gaussian profile fit. We then simultaneously fitted both the primary and the secondary components with HFS line profiles on the original spectra. During this final fit the software CLASS required initial line parameters for both lines: T$_\mathrm{{MB}}$(1,1), v$_\mathrm{{LSR}}$, $\tau$(1,1) and $\Delta$v linewidth. For the primary component the initial parameters were the results of the single HFS line profile fit and for the secondary component v$\mathrm{_{LSR}}$ and $\Delta$v were taken from the HFS line profile fit of the line on the residual, T$\mathrm{_{MB}}$(1,1) from the Gaussian fit and the initial optical depth was 1. None of these parameters were fixed during the simultaneous HFS profile fit. See Fig. \ref{twoline_example} that demonstrates this process. We note that line components were not discernible on any of the NH$_3$(2,2) spectra.

Fitting a single HFS line profile to potentially multiple line components during the first step can be misleading, since to minimize the residual, the fitting process broadens the resulting line. In case the second component can be found on the residual, we can fit the original spectra again, using two components. But if the secondary component is blended sufficiently and does not appear on the residual, a single HFS line profile fit will be able to reproduce the line but give false line parameters. This appears as an uncertainty in our calculations, however, the results are more precise where we identify the secondary component. This way we avoid the detection of a "fake" NH$_3$(1,1) maximum and get a more detailed picture of the velocity structure.

\begin{figure*}[!t]
	\centering
		\includegraphics[scale=0.55]{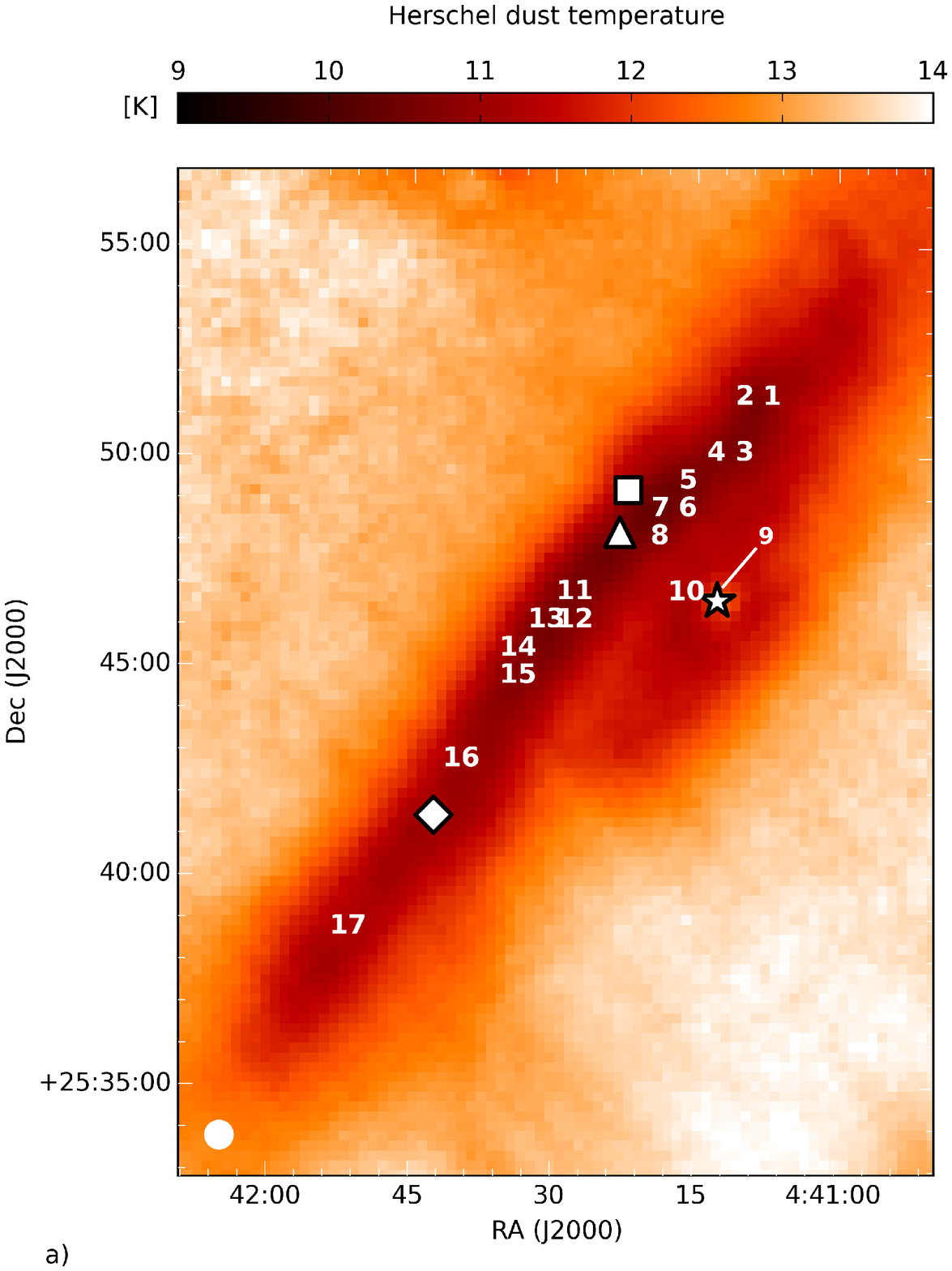}
      	\includegraphics[scale=0.55]{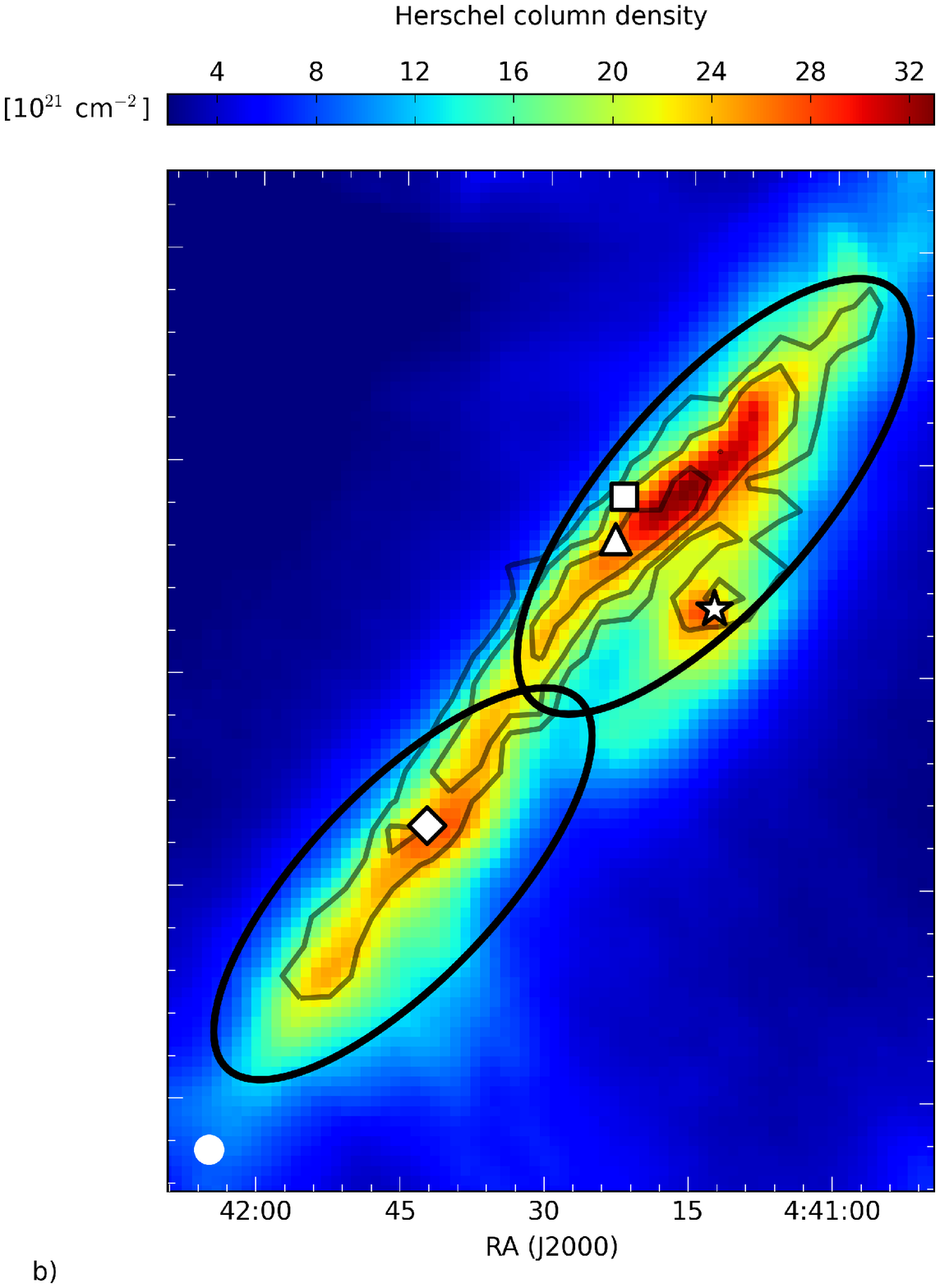}
	\caption{\textbf{a}: \textit{Herschel}-based T$\mathrm{_{dust}}$ map of TMC-1. The numbers mark the positions where high S/N NH$_3$(1,1) and (2,2) spectra were taken and used to calculate the physical parameters of the gas (see Table \ref{amm}). \textbf{b}: \textit{Herschel}-based N(H$_2$) map with the NH$_3$(1,1) main group integrated intensity contours: 0.6, 1.0 and 2.0 Kkms$^{-1}$ (20, 35 and 65\% of the maximum). The secondary line components were subtracted from the spectra. Black ellipses mark the two PGCCs in TMC-1. The positions of TMC-1 CP, TMC-1(NH3), the SO peak and IRAS 04381+2540 are marked with a diamond, a triangle, a square and an asterisk, respectively. HPBW of 40\,$\arcsec$ is shown in the left bottom corner.}
	\label{herschel}
\end{figure*}
We examined the minimum separation of the two line components where it is still possible to identify the weaker line component on the residual. The NH$_3$(1,1) line with the smallest observed linewidth was added to an example spectra while its intensity and their channel separation varied, then we proceeded to search these spectra for the secondary component with our method. With a channel width of 0.08\,kms$^{-1}$ and rms noise in the range of 0.1-0.3\,K, we can identify the secondary line component with a channel separation of 3 channels (0.24\,kms$^{-1}$) only if its intensity is greater than 50\% of the primary component. With a channel separation of 4 channels (0.32\,kms$^{-1}$) the secondary line with 30\% of the primary line component intensity can already be identified. The intensities of the detected secondary components were between 10\% and 40\% of the primary component with velocity separations of about 0.08\,km/s. Since we fitted both components with HFS line profiles and the separation of the two lines is always greater than the HFS linewidths, it is possible to detect close lines in a few cases.

\subsection{Gas parameter calculations from NH$_3$ observations}
\label{ammcalc}
When calculating gas temperatures and densities from the spectra, the primary (or if no secondary component was identified, the only) NH$_3$(1,1) line component was used along with the NH$_3$(2,2) line on the positions where the (2,2) line was detected with S/N\,>\,3. We fitted the primary NH$_3$(1,1) components with Gaussian functions to obtain T$\mathrm{_{MB}}$(1,1) and with HFS line profiles to get $\tau$(1,1), v$\mathrm{_{LSR}}$ and $\Delta$v. The NH$_3$(2,2) lines were fitted with Gaussian functions to get their peak main beam brightness temperatures, T$\mathrm{_{MB}}$(2,2). 

The calculations were done similarly as described by \citet{ho1983}, \citet{ungerechts1986} and \citet{harju1993}. The T$\mathrm{_{rot}}$ rotation temperature of NH$_3$ was calculated with
\begin{equation}
T_{12}=\frac{-41.5}{\ln\left(\frac{-0.282}{\tau(1,1)}\ln\left(1-\frac{T\mathrm{_{MB}}(2,2)}{T\mathrm{_{MB}}(1,1)}\left(1-e^{-\tau(1,1)}\right)\right)\right)}
\end{equation}
The T$\mathrm{_{ex}}$ excitation temperature of the NH$_3$(1,1) transition was derived from
\begin{equation}
T\mathrm{_{MB}}(1,1)=\frac{h\nu_{11}}{k}[F(T\mathrm{_{ex}})-F(T\mathrm{_{bg}})]\left(1-e^{\tau(1,1)}\right) 
\end{equation}
where $\nu_{11}$ is the frequency of the (1,1) transition, T$\mathrm{_{bg}}$ is the cosmic background temperature (2.726\,K), F(T)=1/(e$^{h\nu_{11}/kT}$-1) and we assume a beam filling factor of 1 since the source is a nearby, extended cloud. We calculated the column density of the ammonia molecules in the upper transition level with
\begin{equation}
N\mathrm{_u}=\frac{8\pi\nu_{11}^3}{c^3A}F(T\mathrm{_{ex}})\int\tau(1,1)(v)dv
\end{equation}
where A is the Einstein coefficient of the NH$_3$(1,1) transition (1.7\,$\times$\,10$^{-7}$s$^{-1}$). Assuming a Gaussian line profile and integrating $\tau$ over the main line group we can write the integral as
\begin{equation}
\int\tau(1,1)(v)dv=\frac{\sqrt{\pi}}{2\sqrt{\ln2}}\Delta v\tau(1,1)
\end{equation}

\begin{table*}[hbtp]
	\caption{Results from our NH$_3$ and the \textit{Herschel} observations.}
	\centering
	\resizebox{\textwidth}{!}{
		\begin{tabular}{l c c r c r r r r c r | c c}
			\hline
            \multicolumn{11}{c|}{NH$_3$ observations} & \multicolumn{2}{c}{\textit{Herschel} observations} \\
            \hline
			\multicolumn{1}{c}{N} & \multicolumn{1}{c}{$\Delta\alpha$} & \multicolumn{1}{c}{$\Delta\delta$} & \multicolumn{1}{c}{$\Delta$v} & \multicolumn{1}{c}{$\tau$} & \multicolumn{1}{c}{T$\mathrm{_{kin}}$} & \multicolumn{1}{c}{T$\mathrm{_{ex}}$} & \multicolumn{1}{c}{T$\mathrm{_{rot}}$} & \multicolumn{1}{c}{N$\mathrm{_p}$(NH$_3$)$\mathrm{_{prim}}$} & \multicolumn{1}{c}{N$\mathrm{_p}$(NH$_3$)$\mathrm{_{sec}}$} & \multicolumn{1}{c|}{n(H$_2$)} & \multicolumn{1}{c}{N(H$_2$)} & T$\mathrm{_{dust}}$ \\
			& \multicolumn{1}{c}{[\,$\arcsec$]} & \multicolumn{1}{c}{[\,$\arcsec$]} & \multicolumn{1}{c}{[kms$^{-1}$]} &  & \multicolumn{1}{c}{[K]} & \multicolumn{1}{c}{[K]} & \multicolumn{1}{c}{[K]} & \multicolumn{2}{c}{[10$^{14}$ cm$^{-2}$]} & \multicolumn{1}{c|}{[10$^{4}$ cm$^{3}$]} & \multicolumn{1}{c}{[10$^{22}$ cm$^{-2}$]} & [K] \\
			\hline \hline
1* & -480 &  600 & 0.251 $\pm$ 0.006 & 2.5 $\pm$ 0.2 & 12.9 $\pm$ 1.2 & 5.1 $\pm$ 0.2 & 11.9 $\pm$ 0.9 & 4.73 $\pm$  0.5 & $\le$ 0.5 & 0.7 $\pm$ 0.1 & 2.4 & 11.0 \\
2* & -440 &  600 & 0.253 $\pm$ 0.004 & 2.9 $\pm$ 0.2 & 10.1 $\pm$ 0.8 & 5.1 $\pm$ 0.2 &  9.7 $\pm$ 0.7 & 5.29 $\pm$  0.4 & $\le$ 0.5 & 0.8 $\pm$ 0.2 & 1.4 & 11.7 \\
3* & -440 &  520 & 0.255 $\pm$ 0.006 & 2.9 $\pm$ 0.2 & 11.3 $\pm$ 0.8 & 5.3 $\pm$ 0.2 & 10.7 $\pm$ 0.7 & 5.64 $\pm$  0.5 & $\le$ 0.5 & 0.9 $\pm$ 0.2 & 3.1 & 10.7 \\
4* & -400 &  520 & 0.253 $\pm$ 0.002 & 3.7 $\pm$ 0.1 &  9.0 $\pm$ 0.5 & 5.8 $\pm$ 0.3 &  8.7 $\pm$ 0.4 & 7.63 $\pm$  0.5 & $\le$ 0.5 & 1.6 $\pm$ 0.4 & 2.7 & 10.8 \\
5* & -360 &  480 & 0.251 $\pm$ 0.004 & 5.5 $\pm$ 0.2 &  9.5 $\pm$ 0.6 & 6.3 $\pm$ 0.4 &  9.1 $\pm$ 0.5 & 12.41 $\pm$  1.0 & $\le$ 0.2 & 1.9 $\pm$ 0.5 & 2.9 & 10.6 \\
6 & -360 &  440 & 0.295 $\pm$ 0.008 & 2.8 $\pm$ 0.2 & 11.4 $\pm$ 0.9 & 6.2 $\pm$ 0.3 & 10.8 $\pm$ 0.8 & 7.45 $\pm$  0.8 & ... &  1.4 $\pm$ 0.3 & 3.1 & 10.4 \\
7* & -320 &  440 & 0.332 $\pm$ 0.003 & 4.5 $\pm$ 0.1 &  9.0 $\pm$ 0.4 & 6.7 $\pm$ 0.4 &  8.7 $\pm$ 0.3 & 14.20 $\pm$  1.0 & $\le$ 0.5 & 2.8 $\pm$ 0.9 & 2.9 & 10.5 \\
8 & -320 &  400 & 0.391 $\pm$ 0.015 & 3.4 $\pm$ 0.3 & 10.6 $\pm$ 1.1 & 5.4 $\pm$ 0.3 & 10.1 $\pm$ 0.9 & 10.46 $\pm$  1.2 & ... & 1.0 $\pm$ 0.3 & 2.7 & 10.6 \\
9* & -400 &  320 & 0.278 $\pm$ 0.004 & 1.6 $\pm$ 0.1 & 12.0 $\pm$ 0.8 & 5.5 $\pm$ 0.2 & 11.2 $\pm$ 0.6 & 3.47 $\pm$  0.3 & $\le$ 0.6 & 0.9 $\pm$ 0.1 & 2.4 & 11.3 \\
10* & -360 &  320 & 0.221 $\pm$ 0.012 & 2.8 $\pm$ 0.6 & 13.7 $\pm$ 1.8 & 4.2 $\pm$ 0.2 & 12.6 $\pm$ 1.5 & 3.95 $\pm$  0.9 & $\le$ 0.7 & 0.4 $\pm$ 0.1 & 1.8 & 11.4 \\
11 & -200 &  320 & 0.249 $\pm$ 0.009 & 4.3 $\pm$ 0.4 &  9.2 $\pm$ 0.6 & 6.3 $\pm$ 0.3 &  8.8 $\pm$ 0.4 &  8.88 $\pm$  0.6 &... & 2.1 $\pm$ 0.5 & 2.3 & 10.7 \\
12 & -200 &  280 & 0.308 $\pm$ 0.007 & 2.4 $\pm$ 0.2 & 10.4 $\pm$ 0.9 & 5.0 $\pm$ 0.2 &  9.9 $\pm$ 0.8 & 5.24 $\pm$  0.5 &... & 0.8 $\pm$ 0.2 & 2.4 & 10.6 \\
13 & -160 &  280 & 0.356 $\pm$ 0.006 & 3.3 $\pm$ 0.1 &  9.7 $\pm$ 0.7 & 6.0 $\pm$ 0.3 &  9.3 $\pm$ 0.5 &  9.93 $\pm$  0.7 &... & 1.5 $\pm$ 0.4 & 2.1 & 10.7 \\
14* & -120 &  240 & 0.249 $\pm$ 0.009 & 4.3 $\pm$ 0.4 & 10.5 $\pm$ 1.1 & 5.0 $\pm$ 0.2 & 10.0 $\pm$ 0.9 & 7.70 $\pm$  0.9 & $\le$ 0.3 & 0.8 $\pm$ 0.2 & 2.0 & 10.9 \\
15 & -120 &  200 & 0.288 $\pm$ 0.005 & 2.7 $\pm$ 0.2 & 11.0 $\pm$ 0.7 & 5.7 $\pm$ 0.2 & 10.5 $\pm$ 0.5 & 6.45 $\pm$  0.5 &... & 1.1 $\pm$ 0.2 & 2.4 & 10.8 \\
16 &  -40 &   80 & 0.359 $\pm$ 0.006 & 1.0 $\pm$ 0.1 & 11.0 $\pm$ 1.0 & 5.9 $\pm$ 0.2 & 10.4 $\pm$ 0.8 & 2.93 $\pm$  0.2 &... &1.2 $\pm$ 0.3 & 2.1 & 11.1 \\
17* &  120 & -160 & 0.262 $\pm$ 0.005 & 1.6 $\pm$ 0.2 & 10.1 $\pm$ 0.9 & 4.7 $\pm$ 0.1 &  9.7 $\pm$ 0.8 & 2.84 $\pm$  0.3 & $\le$ 0.3 & 0.7 $\pm$ 0.1 & 2.4 & 11.3 \\ 
		\hline
		\end{tabular}}
\tablefoot{Numbers in the first column refer to the marked positions in Fig. \ref{herschel}a and asterisk marks where a secondary line component was found.}
	\label{amm}
\end{table*}
The N(NH$_3$(1,1)) total column density of the NH$_3$ molecules in the (1,1) state is
\begin{equation}
\label{n11}
N(NH_3((1,1))=N\mathrm{_u}(1,1)+N\mathrm{_l}(1,1)=N\mathrm{_u}(1,1)(1+e^{h\nu_{11}/kT\mathrm{_{ex}}})
\end{equation}
Ammonia has two distinct species, the ortho-NH$_3$ (K\,=\,3$n$) and the para-NH$_3$ (K\,$\neq$\,3$n$), which arise from different relative orientations of the three hydrogen spins. Since the higher energy levels are by orders of magnitude less populated on temperatures around 10\,K, we may estimate the N$\mathrm{_p}$(NH$_3$) para-NH$_3$ column density from the observed (1,1) and (2,2) transitions with
\begin{equation}
\label{ntotal}
N\mathrm{_p}(NH_3)=N(NH_3(1,1))\left(1+\frac{5}{3}e^{-41.5/T_{12}}\right)
\end{equation}
The T$\mathrm{_{kin}}$ kinetic temperature of the gas was calculated with the half-empiric equation from \citet{tafalla2004}:
\begin{equation}
T\mathrm{_{kin}}=\frac{T_{12}}{1-\frac{T_{12}}{42}\ln\left(1+1.1e^{-16/T_{12}}\right)}
\end{equation}
The n(H$_2$) local molecular hydrogen volume density was estimated using the equation from \citet{ho1983}:
\begin{equation}
n(H_2)=\frac{A}{C}\frac{F(T\mathrm{_{ex}})-F(T\mathrm{_{bg}})}{F(T\mathrm{_{kin}})-F(T\mathrm{_{ex}})}[1+F(T\mathrm{_{kin}})]
\end{equation}
where C is the collisional de-excitation rate (8.5\,$\times$\,10$^{-11}$ cm$^3$s$^{-1}$) from \citet{danby1988}.

We calculated N$\mathrm{_p}$(NH$_3$) upper limits from the secondary NH$_3$(1,1) line components at each of the 31 positions where those were detected (see Fig. \ref{vel} and Appendix \ref{ammdens}). Based on the HFS line profile fits, on 82 positions the $\tau$(1,1) optical depth had a relative error less than 50\% (median relative error was 17\%), so from these high S/N spectra the N(NH$_3$(1,1)) column density could be estimated.

\subsection{\textit{Herschel} observations and data reduction}
\label{herscheldata}

The Taurus region has been observed as part of the \textit{Herschel Gould Belt Survey} \citep{andre2010, kirk2013}. The {SPIRE} instrument (Spectral and Photometric Imaging Receiver) used three arrays of bolometers to observe wavelengths centered at 250, 350 and 500 $\mu$m with bandwidths of 33\% each, effectively covering the 208-583\,$\mu$m range \citep{griffin2010}. The FWHM beam sizes were 17.6\,$\arcsec$, 23.9\,$\arcsec$ and 35.2\,$\arcsec$ for the three bands respectively and the field-of-view was 4\,$\times$\,8\,$\arcmin$. \textit{PACS} (Photodetector Array Camera and Spectrometer) observed wavelengths from 60 to 210\,$\mu$m with two bolometer arrays, simultaneously with the 125-210\,$\mu$m band (red) and with either the 60-85 or the 85-125\,$\mu$m band (blue) \citep{poglitsch2010}. The blue channels had 32\,$\times$\,64 pixel arrays and the red channel had a 16\,$\times$\,32 pixels array. Both channels covered a field-of view of 1.75\,$\times$\,3.5\,$\arcmin$ with full beam spacing in each band. 

TMC-1, as a part of HCL 2, was mapped using the fast-scanning \textit{SPIRE/PACS} Parallel Mode in two orthogonal scan directions. The \textit{Herschel} observation IDs of the data are 1342202252 and 1342202253. The data presented in this paper were processed with HIPE\footnote{Herschel Interactive Processing Environment, http://herschel.esac.esa.int/hipe/} 12.1.0. The level-0.5 timelines were calibrated and converted to physical units separately and joined in the level-1 stage. The destriper module was applied to remove the observation baseline. Then the sourceExtractorTimeline and the sourceExtractorSussextractor tasks were used to search for point sources in the 
data with a given FWHM and to determine their fluxes. The point sources then were subtracted with the sourceSubtractor task from the timeline data of all three wavelengths. See the HIPE Help System\footnote{http://herschel.esac.esa.int/hcss-doc-14.0/} for the description of the applied tasks. We located 4 point sources on the 350\,$\mu$m timeline, one of them, \object{IRAS 04381+2540} is also seen on 250 and 500\,$\mu$m. After point source subtraction the maps were combined using the naive map-making method. Absolute calibrated maps were produced with the ZeroPointCorrection task which calculates the absolute offsets by cross-correlating with \textit{Planck HFI} 545 and 857\,GHz data. The calibration accuracy of the \textit{SPIRE} data is estimated to be 7\%. After convolution to 40\,$\arcsec$ common resolution the maps were co-aligned on the 500\,$\mu$m pixel-grid with 14\,$\arcsec$ pixel width and each pixel was colour-corrected.

\subsection{Calculations from \textit{Herschel SPIRE} data}
\label{herschelcalc}

We used the method by \citet{juvela2012b} to derive hydrogen column density from the \textit{SPIRE} maps. We fitted the spectral energy distribution (SED) of each pixel with $B_{\nu}(T\mathrm{_{dust}})\nu^{\beta}$ where B$_{\nu}$(T) is the Planck-function for a black-body with colour temperature $T\mathrm{_{dust}}$ at $\nu$ frequency. The value of $\beta$ varies between 1.8 to 2.2 in the cold dense ISM, anti-correlates with temperature and correlates with column density and galactic position \citep{juvela2015b}, see Appendix \ref{beta} for further information. In our calculations the SED was fitted to all three \textit{SPIRE} data points with a fixed $\beta$\,=\,2 spectral index and returned T$\mathrm{_{dust}}$ for each pixel.

The N(H$_2$) molecular hydrogen column density was calculated as:
\begin{equation}
N(H_2)=\frac{I_{\nu}}{B_{\nu}(T\mathrm{_{dust}})\kappa m\mathrm{_H\mu}}
\end{equation}
where $\mu$\,=\,2.33 the particle mass per hydrogen molecule, m$\mathrm{_H}$ is the atomic mass unit, $\kappa$ is the dust opacity from the formula 0.1\,cm$^2$/g($\nu$/1000 GHz)$^{\beta}$ for high-density environments \citep{beckwith1990, planck2011c} and I$_{\nu}$ is the intensity at $\nu$\,=\,856\,GHz ($\lambda$\,=\,350 $\mu$m). The 350\,$\mu$m \textit{Herschel} band overlaps with the \textit{Planck} 857\,GHz band which was used for zero-point correction. This way we minimize the error that originates from the absolute calibration.
\begin{figure}[!tp]
	\centering
		\includegraphics[angle=-90, trim=0 1cm 0 0, width=1.1\linewidth]{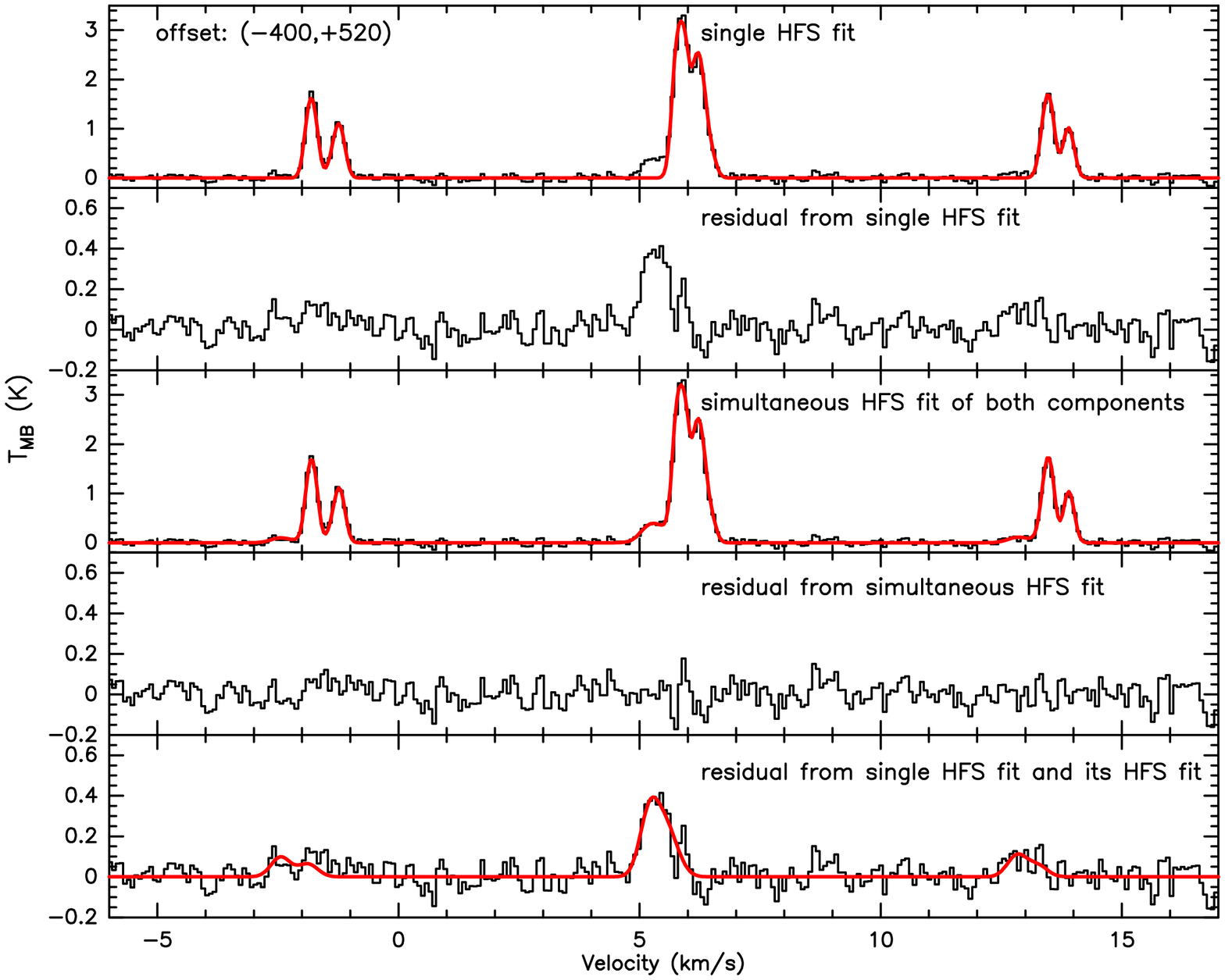}
   		\includegraphics[angle=-90, trim=1cm 1cm 7.5cm 0, width=1.1\linewidth]{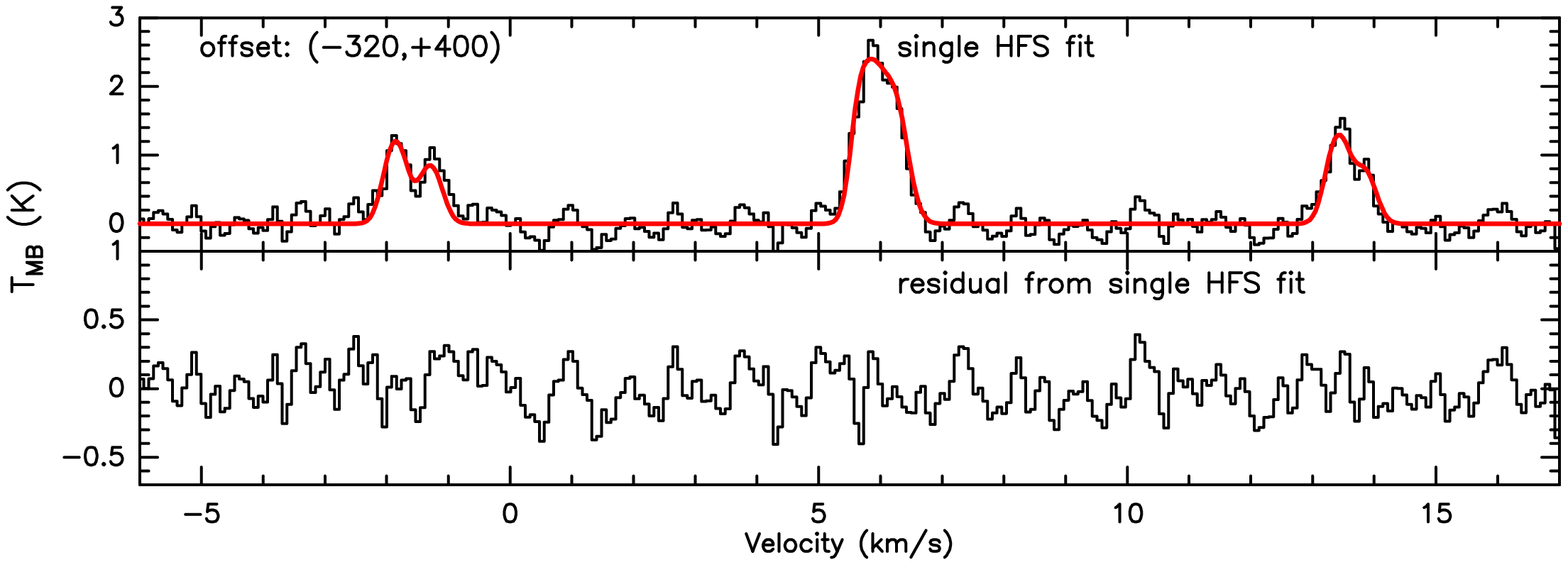}
		\caption{\textbf{Top:} Multiple HFS line profile fitting of NH$_3$(1,1) on the (-400,520) offset. The spectrum shows one of the brightest secondary line components and gives the lowest T$\mathrm{_{kin}}$ value, see position \#4 in Table \ref{amm}. Box 1: single HFS line profile fit; box 2: residual from the single HFS fit; box 3: simultaneous HFS profile fitting of both components; box 4: the residual from the simultaneous HFS fit; box 5: the residual from the single HFS fit with the HFS fit of the secondary component shown in box 4. \textbf{Bottom:} Single NH$_3$(1,1) HFS line profile fit on position \#8 and the residual after subtracting the fit. This position has the highest HFS linewidth from the spectra in Table \ref{amm}. We note the different scales for the (1,1) and (2,2) lines.}
		\label{twoline_example}
\end{figure}

The N(H$_2$) distribution calculated this way does not exclude the contribution of structures in the foreground or the background of HCL 2. In order to correct for this, the extent of HCL 2 was marked with the A$\mathrm{_V}$=1 magnitude contour \citep[9.4\,$\times$\,10$^{20}$ cm$^{-2}$, where interstellar hydrogen starts to become molecular;][]{bohlin1978} on the N(H$_2$) map derived as written above. Then the average 250, 350 and 500\,$\mu$m intensity at this contour was subtracted from the corresponding \textit{SPIRE} 250, 350 and 500\,$\mu$m maps. The final T$\mathrm{_{dust}}$ and N(H$_2$) maps were calculated from these background subtracted intensity images. Other methods of background subtraction are evaluated in Appendix \ref{bground}.

\citet{pagani2015} concluded that a single-temperature SED fitting we use here may miss some of the cold dust in dense cores, thus we probably slightly underestimate the total column density. With only three \textit{SPIRE} data points, we cannot fit multi-temperature SEDs to model an even colder component but this can be checked with independent methods, such as star-reddening. \citet{malinen2012} compared FIR emission based and NIR reddening based N(H$_2$) values in HCL 2, and they found that the two methods are in good agreement. A significant source of uncertainty in the N(H$_2$) calculation is the error in the surface brightness, as shown by \citet{juvela2012b} using Monte Carlo simulations. Assuming a 13\% uncertainty in the \textit{Herschel} intensities an error of less than 1.5\,K appears in regions with temperatures below 16\,K. This corresponds to an error in N(H$_2$) of a few times 10$^{21}$\,cm$^{-2}$ in the outer region of the filament and around 1-2\,$\times$\,10$^{22}$ cm$^{-2}$ in the densest parts.
\section{Results}
\label{results}

\subsection{T$\mathrm{_{dust}}$ and N(H$_2$) distribution from \textit{Herschel} data}
\label{herschelres}
The \textit{Herschel}-based N(H$_2$) map of HCL 2 is shown in Fig. \ref{hcl22} where the PGCCs are represented by numbered ellipses. All PGCCs with flux quality of 1 (flux density estimates S/N\,>\,1 in all bands) and 2 (flux density estimates S/N\,>\,1 only in the 857, 545 and 353\,GHz \textit{Planck} bands) are plotted on the figure. Table~\ref{pgcc2} contains the calculated parameters of these PGCCs, where the columns are: (1) number, also in Fig. \ref{hcl22}; (2) PGCC ID; (3,4) equatorial coordinates; (5,6) T$\mathrm{_{dust,min}}$ minimum and T$\mathrm{_{dust,ave}}$ average colour temperature in the clump; (7,8) N(H$_2$)$\mathrm{_{max}}$ maximum and N(H$_2$)$\mathrm{_{ave}}$ average column density in the clump; (9) other IDs of the clump. PGGCs in TMC-1, \object{TMC-1C} and HCL 2B are similarly cold with T$\mathrm{_{dust,min}}$\,$\leq$\,11\,K and dense with N(H$_2$)$\mathrm{_{ave}}$\,$\geq$\,7\,$\times$\,10$^{21}$\,cm$^{-2}$.
\begin{figure*}[!t]
	\centering
		\includegraphics[scale=0.55]{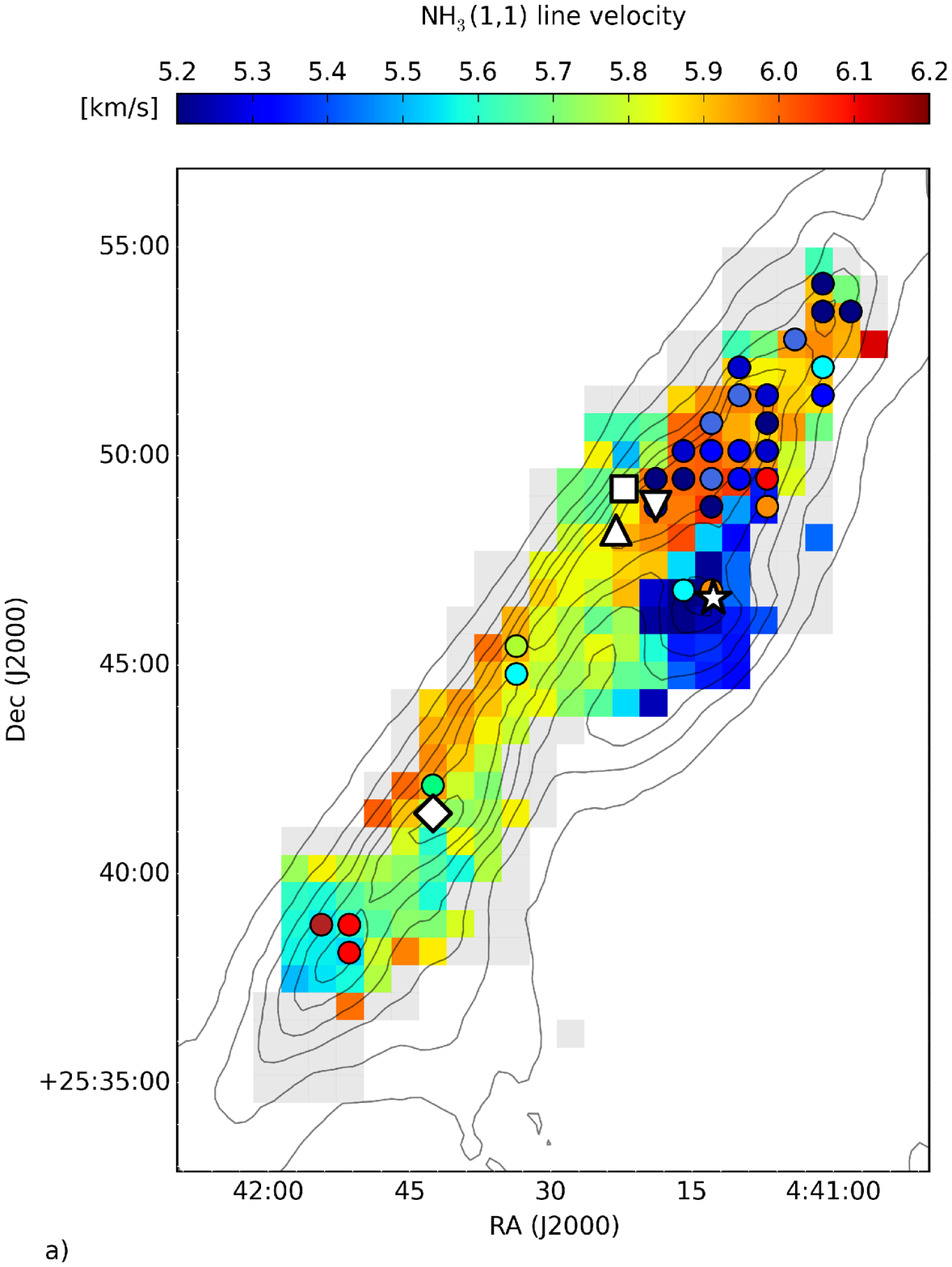}
       \includegraphics[scale=0.55]{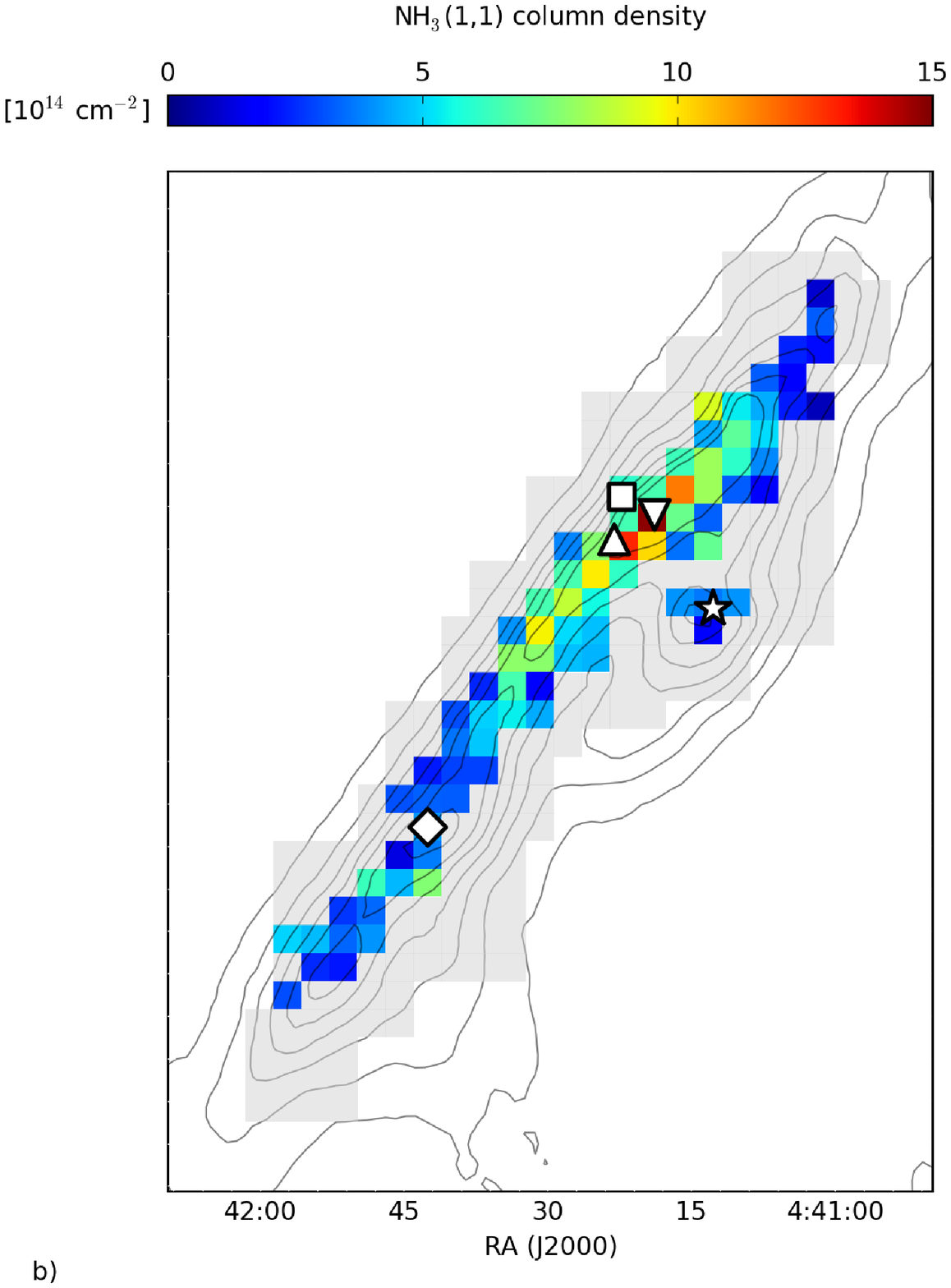}
	\caption{\textbf{a}: v$\mathrm{_{LSR}}$ distribution of the main (or only) line component of the NH$_3$(1,1) lines with S/N\,>\,3. Circles mark the positions where a secondary component was found and their color corresponds to the velocity of this component. \textbf{b}: N(NH$_3$(1,1)) distribution map calculated on positions where the $\tau$ optical depth from single HFS profile fit had an error less than 50\%. The grey area shows the total extension of the observed NH$_3$ map and the grey contours are the \textit{Herschel} N(H$_2$): 0.7, 1.0, 1.3, 1.6, 2, 2.3, 2.6\,$\times$\,10$^{22}$\,cm$^{-2}$ (from 20 to 80\% of the maximum by 10\%). The positions of TMC-1 CP, TMC-1(NH3), the SO peak and IRAS 04381+2540 are marked as before and the position of the our NH$_3$ peak is marked with an upside down triangle.}
	\label{vel}
\end{figure*} 

The \textit{Herschel} T$\mathrm{_{dust}}$ map of TMC-1 is shown in Fig. \ref{herschel}a with the positions of high NH$_3$(2,2) S/N spectra, where temperatures and densities from the NH$_3$ data could be calculated (see Section \ref{nh3res}). T$\mathrm{_{dust}}$ has a minimum of 10.4\,K in the northern, densest part and is 11\,K at its southern end. Lower density outskirts are a few K warmer.

The \textit{Herschel} N(H$_2$) map of TMC-1 is shown in Fig. \ref{herschel}b with the NH$_3$(1,1) main group integrated intensity contours overplotted (the secondary line components were already subtracted from the spectra). PGCC G174.20-13.44 and PGCC G174.40-13.45 are marked by ellipses and the positions of TMC-1 CP, TMC-1(NH3), the SO peak and IRAS 04381+2540 are also marked. The map shows TMC-1 as a narrow ridge with an average diameter of 3-4\,$\arcmin$ (0.12-0.17\,pc). Three local N(H$_2$) maxima appear in TMC-1. Two very elongated ones are along the ridge with peaks 2\,$\arcmin$ north-west from TMC-1(NH3) with N(H$_2$)\,=\,3.3$\times$\,10$^{22}$cm$^{-2}$ and $\approx$\,30\,$\arcsec$ south-west from TMC-1 CP with N(H$_2$)\,=\,2.7\,$\times$\,10$^{22}$cm$^{-2}$. The third local N(H$_2$) maximum is at $\approx$\,18\,$\arcsec$ south-east from the position of the Class I protostar IRAS 04381+2540 with N(H$_2$)\,=\,2.8\,$\times$\,10$^{22}$cm$^{-2}$. The PGCCs ellipses coincide with the northern and the southern ones. The cloud parts can not be simply separated by the half power contours, since the column density remains above 10$^{22}$\,cm$^{-2}$ in the central ridge along the filament. We cannot achieve a robust separation of sub-structures inside TMC-1 (see Appendix \ref{clfind}) using the N(H$_2$) distribution only.

\subsection{Results from NH$_3$ measurements}
\label{nh3res}

The NH$_3$(1,1) integrated intensity contours coincide well with the N(H$_2$) distribution, as seen in Fig. \ref{herschel}b. The integrated intensity has a peak at the densest, coldest part of the filament, $\approx$1\,$\arcmin$ to south-east of TMC-1(NH3), then the line intensities decrease gradually towards the southern edge, while N(H$_2$) remains over 10$^{22}$ cm$^{-2}$ even at the southern end of the ridge.

The measured NH$_3$(1,1) primary component HFS linewidths are in the range of 0.15-1.0\,kms$^{-1}$, with an average of $\Delta$v of 0.39\,kms$^{-1}$. Due to the small linewidths and with our high velocity resolution, subregions could be separated by their velocities and multiple velocity components could be identified. The spatial distribution of velocities of the NH$_3$(1,1) lines with S/N\,>\,3 on 171 positions are plotted in Fig. \ref{vel}a. The area marked by these positions is above the \textit{Herschel}-based 10$^{22}$\,cm$^{-2}$ column density contour. At the northern part of the ridge the velocities are around 5.9-6.0\,kms$^{-1}$ and most spectra show a secondary line component with velocities around 5.2-5.5\,kms$^{-1}$. South-west of the main ridge, near IRAS 04381+2540, a Class\,I young stellar source \citep{terebey1980} the primary line velocities are lower (v$\mathrm{_{LSR}}$\,$\approx$\,5.2\,kms$^{-1}$). The secondary line components in this same region have v$\mathrm{_{LSR}}$\,$\approx$\,5.9\,kms$^{-1}$, except one with 5.6\,kms$^{-1}$. There is a 4.9\,kms$^{-1}$pc$^{-1}$ velocity gradient seen in the middle part (25:41:00\,<\,$\delta$(2000)\,<\,25:47:00). The primary line component velocities decrease southward from the mid-region of the ridge from 5.9 to 5.5 km$^{-1}$, while the secondary from the HFS line profile fit was S/N\,>\,3.

Fig. \ref{vel}b shows that N(NH$_3$(1,1)) was calculated on 82 positions that are all above the 1.3\,$\times$\,10$^{22}$\,cm$^{-2}$ \textit{Herschel} contour. The N(NH$_3$(1,1)) distribution follows well the \textit{Herschel}-based N(H$_2$) and reaches 1.5\,$\times$\,10$^{14}$\,cm$^{-2}$ even around the CP peak. Its maximum (1.4\,$\times$\,10$^{15}$\,cm$^{-2}$) 
is at the offset of (-320,440), with absolute coordinates of $\alpha$(2000)\,=\,04:41:13, $\delta$(2000)\,=\,25:48:47 (position 7 in Fig. \ref{herschel}b). That
is $\approx$\,1$\arcmin$ to north-west of the position which was formerly known as the ammonia peak, i.e. TMC-1(NH3).

The NH$_3$(2,2) line was detected with S/N\,>\,3 in as many as 17 spectra, we mark the corresponding locations in TMC-1 as position \#$s$ where the number $s$ runs from 1 to 17 (see Fig. \ref{herschel}a). Parameters at these positions were calculated according to Section \ref{ammcalc} and the results are listed in Table \ref{amm} where the columns are: (1) number of position in Fig. \ref{herschel}, asterisk marks where a secondary line component was found and subtracted before the calculations; (2,3) right ascension and declination offsets; (4) linewidth from HFS line profile fit; (5) optical depth; (6) kinetic temperature; (7) excitation temperature; (8) rotation temperature; (9,10) NH$_3$ column density from the primary and the secondary components respectively; (11) H$_2$-volume density; (12,13) \textit{Herschel} H$_2$-column density and colour temperature. Numbers in the first column refer to the marked positions in Fig. \ref{herschel}. The derived T$\mathrm{_{kin}}$ values vary between 9 and 13\,K with the exception of position \#10 with 13.7\,K. The NH$_3$ column density is 2.8\,$\times$\,10$^{14}$\,<\,N$\mathrm{_p}$(NH$_3$)\,<\,1.4\,$\times$\,10$^{15}$ cm$^{-2}$ and the H$_2$ volume density is around 0.4\,$\times$\,10$^{4}$\,<\,n(H$_2$)\,<\,2.8\,$\times$\,10$^{4}$ cm$^{-3}$. The N$\mathrm{_p}$(NH$_3$) distribution generally follows the \textit{Herschel}-based N(H$_2$) distribution along the ridge, although their linear correlation is not very high (the correlation coefficient is p\,=\,0.4). The N(NH$_3$(1,1)) maximum coincides with the N$\mathrm{_p}$(NH$_3$)\,=\,N$\mathrm{_p}$(NH$_3$)$\mathrm{_{prim}}$\,+\,N$\mathrm{_p}$(NH$_3$)$\mathrm{_{sec}}$ maximum on position \#7 and it should be considered as the new NH$_3$ maximum of TMC-1.

We calculated the NH$_3$ relative abundance at our NH$_3$ peak and close to TMC-1 CP using the derived N(H$_2$) and N$\mathrm{_p}$(NH$_3$) values in Table \ref{amm}. This gives 5\,$\times$\,10$^{-8}$ and 1.4\,$\times$\,10$^{-8}$ respectively, which are consistent with the estimates of \citet{harju1993} for dense cores in Orion and Cepheus (1-5\,$\times$\,10$^{-8}$) and with the chemical modeling of \citet{suzuki1992} (around 10$^{-8}$). Generally we find higher abundances in the northern and mid-region of TMC-1 and lower values around the YSO and in the southern end.

The total velocity dispersion was calculated from the primary NH$_3$(1,1) component of the 17 high S/N spectra with
\begin{equation} \label{totalsig}
\sigma\mathrm{_{total}}=\frac{\Delta v}{\sqrt{8ln(2)}}
\end{equation}
where $\Delta$v is the HFS linewidth. The thermal component of the velocity dispersion is
\begin{equation}
\label{thsig}
\sigma\mathrm{_{th}}=\sqrt{\frac{k\mathrm{_B}T\mathrm{_{kin}}}{\mu_\mathrm{{NH_3}}m\mathrm{_H}}} 
\end{equation}
where $\mu\mathrm{_{NH_3}}$\,=\,17 is the ammonia molar mass and m$\mathrm{_H}$ is the atomic mass unit. The non-thermal (turbulent) component was then derived by
\begin{equation}
\label{nthsig}
\sigma\mathrm{_{nth}}^2=\sigma\mathrm{_{total}}^2-\sigma\mathrm{_{th}}^2 
\end{equation}
The thermal component is around 0.066-0.08\,kms$^{-1}$ in the 17 spectra and the non-thermal component is higher than this in almost every case, except on positions \#1 and \#10. The highest turbulence is observed on positions \#8, \#13 and \#16.

\section{Discussion}

\subsection{Density, temperature and velocity structure}
\label{densvel}

We obtained the most extensive NH$_3$ map of TMC-1 so far where both the (1,1) and (2,2) transitions were measured with a low noise level and high velocity resolution. Our observations confirm the variation in the NH$_3$(1,1) line intensity in TMC-1 that was described by several studies \citep{little1979, olano1988, hirahara1992, pratap1997}. Our average HFS linewidths agree with the values of \citet{toelle1981}, although we mapped a significantly larger area.

\citet{olano1988} measured NH$_3$(1,1) with similar beam size, velocity resolution and S/N in the high-density part of TMC-1. They also observed the velocity difference between the main part of the ridge and its southern end, and a considerably different velocity for the region around IRAS 04381+2540 (see their Fig. 9.). They pointed out a velocity gradient across the ridge as well. We could detect further velocity variations because our mapping extends further along and across the ridge compared to theirs.

The T$\mathrm{_{ex}}$ values we derived agree well with the 4-7\,K range given by \citet{toelle1981}. Their observations do not completely overlap with ours because they measured positions with a spacing of 45\,$\arcsec$, but the difference between their values and ours on positions less than 35\,$\arcsec$ apart is less than 25\%, relative to their values. They calculated T$\mathrm{_{rot}}$ on two positions which are relatively close ($\approx$\,40\,$\arcsec$ and $\approx$\,5\,$\arcsec$) to our positions \#11 and \#15: their values are 9.5\,$\pm$\,1\,K and 10.3\,$\pm$\,2\,K and we derived 8.8\,K and 10.5\,K, respectively. \citet{gaida1984} repeated NH$_3$(1,1) and (2,2) measurements on the positions of \citet{toelle1981} along the main ridge of TMC-1 and made two cuts perpendicular to the major axis as well. They calculated both NH$_3$ column densities and kinetic temperatures on three positions that are not farther than 40\,$\arcsec$ from our position \#11, \#13 and \#15. Their T$\mathrm{_{kin}}$, T$\mathrm{_{rot}}$ results are in agreement with ours but their N(NH$_3$(1,1)) values are consistently 4-5 times higher than ours and N(NH$_3$) only agrees well on the higher density positions. Their v$\mathrm{_{LSR}}$ and n(H$_2$) distribution along the main ridge also shows similar values as measured by us. Our T$\mathrm{_{ex}}$ values are somewhat lower than the ones by \citet{olano1988} (6-9\,K), which they show on a plot along the TMC-1 major axis. Their sample does not overlap with ours either because of the different (0,0) offset coordinates they used. 

\citet{pratap1997} and \citet{hirahara1992} derived a factor of 5-6 higher H$_2$-volume densities from HC$_3$N and C$^{34}$S spectra than our results. Our NH$_3$ measurements may not trace the densest gas in TMC-1 because, as it was shown by \citet{bergin2002} and \citet{pagani2005}, NH$_3$ is best used as a tracer at lower than 5\,$\times$\,10$^{5}$\,cm$^{-3}$ densities. Also because of the low critical density (around 2\,$\times$\,10$^3$\,cm$^{-3}$) of the (1,1) line, it may have a substantial contribution from external, warmer layers as well \citep{pagani2007}. Our results do not show significant differences in volume densities along the ridge, while the observations by \citet{hirahara1992} imply a gradient of a factor of 10 in n(H$_2$) between TMC-1(NH3) and TMC-1 CP. The NH$_3$ densities calculated by \citet{toelle1981} at TMC-1(NH3) are consistent with our results but their values are a factor of 2-3 higher at TMC-1CP. The results by \citet{pratap1997} are consistent with ours at TMC-1 CP but a factor of 2-3 lower than our values in the northern region.
 
\citet{nutter2008} presented temperature and density profile modeling for TMC-1 based on \textit{SCUBA, IRAS} and \textit{Spitzer} continuum data. Their best fit was given by an inner colder filament (8\,K) and an outer layer (12\,K) and resulted in n(H$_2$) values in the range of 0.7-1.75\,$\times$\,10$^5$ cm$^{-3}$. They also calculated N(H$_2$) from the 850\,$\mu$m \textit{SCUBA} map which yielded 3.8\,$\times$\,10$^{22}$ and 4.4\,$\times$\,10$^{21}$ cm$^{-2}$ for the central part and the outer layer respectively. \citet{malinen2012} used \textit{2MASS} and \textit{Herschel} maps to derive the properties for TMC-1. Their results are consistent with the findings of \citet{nutter2008} and their N(H$_2$) peak is around 3.4\,$\times$\,10$^{22}$ cm$^{-2}$. This is our derived N(H$_2$) maximum as well. We note that our calculations are based on point source subtracted \textit{Herschel} images. T$\mathrm{_{dust}}$ values derived by us for the outer part of TMC-1 are consistent with the 12\,K from \citet{nutter2008} but we measure 10\,K in the inner region instead of 8\,K. T$\mathrm{_{kin}}$ derived from NH$_3$ and T$\mathrm{_{dust}}$ derived from the \textit{Herschel} measurements vary similarly but do not agree well, since the cooling-heating processes are different for gas and dust and the coupling of the two phases are not guaranteed in regions with n(H$_2$)\,=\,10$^{4}$ cm$^{-3}$ \citep{goldsmith2001}.
\begin{table}[t]
\caption{Results from the \textit{NbClust} method.}
	\centering
		\begin{tabular}{lccccc}
			\hline
			m & N & I & C$\mathrm{_m}$ & C$\mathrm{_{m,w}}$ & Figure \\
			\hline \hline
			k-means & 4 & 11 & 0.0054 & 0.25 & Fig. \ref{wardclumping} \\
			\hline
            McQuitty\tablefootmark{*} & 15 & 6 & 0.0029 & 0.048 & Fig. \ref{clumping2}a\\
		    Ward & 4 & 7 & 0.0059 & 0.38 & Fig. \ref{clumping2}b \\
			Ward2 & 4 & 8 & 0.0069 & 0.38 & Fig. \ref{clumping2}c\\
            complete & 4 & 9 & 0.0086 & 0.44 & Fig. \ref{clumping2}d\\
			average & 3 & 7 & 0.0123 & 0.78 & Fig. \ref{clumping2}e\\
            median & 2 & 7 & 0.0126 & 2.16 & Fig. \ref{clumping2}f\\
            centroid & 2 & 9 & 0.0129 & 2.22 & Fig. \ref{clumping2}g\\
			single & 2 & 8 & 0.0131 & 2.26 & Fig. \ref{clumping2}h\\
			\hline
		\end{tabular}
	\tablefoot{The columns are: (1) name of the clustering method; (2) ideal number of clusters; (3) number of indices proposing N; (4,5) sum (equation \ref{sumofvar}) and weighted sum (equation \ref{wsumofvar}) of the within-cluster variances; (6) reference to the figure showing the clusters projected to the plane of the sky. \\
       \tablefoottext{*}{too small fragments relative to the beam size}}

	\label{nbres}
\end{table}
\begin{figure}[!t]
    \includegraphics[width=\linewidth]{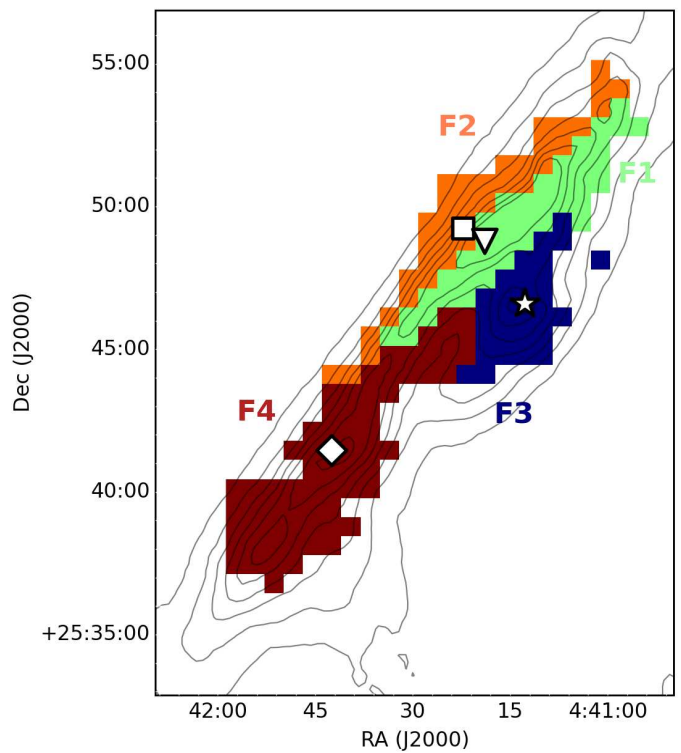}
	\caption{The resulting clusters from the k-means clustering method projected back to the plane of the sky. The grey contours are the \textit{Herschel} N(H$_2$): 0.7, 1.0, 1.3, 1.6, 2, 2.3, 2.6\,$\times$\,10$^{22}$\,cm$^{-2}$ (from 20 to 80\% of the maximum by 10\%). The positions of TMC-1 CP, IRAS 04381+2540, the SO peak and our NH$_3$ maximum are marked as before.}
	\label{wardclumping}
\end{figure}

\begin{table*}[ht]
\caption{Physical parameters of the sub-filaments. See Fig. \ref{wardclumping}}
	\centering
    	\resizebox{\textwidth}{!}{
		\begin{tabular}{r c c c c c r c r c c c c c}
			\hline
			\multicolumn{1}{c}{Sub-} & \multicolumn{1}{c}{$\alpha$(2000)} & \multicolumn{1}{c}{$\delta$(2000)} & \multicolumn{1}{c}{v$\mathrm{_{LSR,ave}}$} & \multicolumn{1}{c}{N(H$_2$)$\mathrm{_{ave}}$} & \multicolumn{1}{c}{T$\mathrm{_{dust,ave}}$} & \multicolumn{1}{c}{T$\mathrm{_{kin,ave}}$} & \multicolumn{1}{c}{$\Delta$v$\mathrm{_{ave}}$} & \multicolumn{1}{c}{L} & \multicolumn{1}{c}{r} & \multicolumn{1}{c}{M$\mathrm{_{lin,obs}}$} & \multicolumn{1}{c}{M$\mathrm{_{lin,vir}}$} & \multicolumn{1}{c}{$\sigma\mathrm{_{th,ave}}$} & \multicolumn{1}{c}{$\sigma\mathrm{_{nth,ave}}$}  \\
			\multicolumn{1}{c}{filament} & [deg] & [deg] & [kms$^{-1}$] & \multicolumn{1}{c}{[10$^{22}$ cm$^{-2}$]} & \multicolumn{1}{c}{[K]} & \multicolumn{1}{c}{[K]} & \multicolumn{1}{c}{[kms$^{-1}$]} & \multicolumn{1}{c}{[pc]} & \multicolumn{1}{c}{[pc]} & \multicolumn{1}{c}{[M$_{\sun}$/pc]} & \multicolumn{1}{c}{[M$_{\sun}$/pc]} & \multicolumn{1}{c}{[kms$^{-1}$]} & \multicolumn{1}{c}{[kms$^{-1}$]} \\
			\hline
 TMC-1F1 & 70.3048 & 25.8235 & 5.93 &  2.2 & 11.1 & 10.3 & 0.359 & 0.5 & 0.14 & 32.5 & 25.4 & 0.071 & 0.135 \\
TMC-1F2 & 70.3341 & 25.8302 & 5.90 &  1.8 & 11.4 &  9.9 & 0.383 & 0.6 & 0.13 & 19.6 & 26.3 & 0.069 & 0.147 \\
TMC-1F3 & 70.3026 & 25.7748 & 5.28 &  1.7 & 11.6 & 12.9 & 0.407 &  0.3 & 0.15 & 28.9 & 32.2 & 0.079 & 0.154 \\
TMC-1F4 & 70.4220 & 25.6879 & 5.77 &  1.8 & 11.6 & 10.4 & 0.527 & 0.5 & 0.19 & 45.9 & 38.1 & 0.071 & 0.212 \\
\hline
		\end{tabular}}
	\tablefoot{The columns are: (1) number in Fig. \ref{wardclumping}; (2,3) coordinates of the sub-filament center; (4) average v$\mathrm{_{LSR}}$; (5,6) average \textit{Herschel} H$_2$ column density and colour temperature in the sub-filament; (7,8) average kinetic temperature and linewidth in the sub-filament; (9,10) length and radius of the clump; (11,12) linear mass and virial mass per unit length of the sub-filament; (13,14) average thermal and non-thermal velocity dispersion in the sub-filament.}
	\label{clusters}
\end{table*}

\subsection{Sub-structures in TMC-1}
\label{clustering}

\subsubsection{Clustering methods and their results}
\label{methods}

The complex velocity structure and the appearance of secondary velocity components on several spectra suggest different cloud parts being present in TMC-1. Since the filament is not necessarily one coherent structure and it could be interpreted as a chain of clumps, or even a complex superposition of filaments containing smaller-scale cores, we apply several types of clustering methods to reveal its morphology.

Clump-finding in the NH$_3$ PPV datacube with methods like the 3D version of \textsc{clumpfind}, \textsc{gaussclump} or the dendogram technique (see Appendix \ref{clfind}) is problematic in the case of TMC-1, since NH$_3$ abundance can be position-dependent in the cloud. Moreover, NH$_3$ alone is not a good tracer for clumps because of high optical depth effects like self-absorption, which affects the line profiles. Searching for clumps in the \textit{Herschel}-based N(H$_2$) distribution is not enough either, since N(H$_2$) changes smoothly along the ridge. As we can see from a quick decomposition of the filament with \textsc{clumpfind} in Appendix \ref{clfind}, we do not get a robust result: the extent and shape of the derived clumps strongly depend on the initial parameters given to the clump finding algorithm. Also, this way we ignore the velocity information derived from molecular emission that traces the relative motions of the gas and can help identifying structures in the line of sight.

We expect that the presence of a clump changes the N(H$_2$) distribution in its direction. We assume that inside a clump N(H$_2$) changes continuously, velocities scatter around a certain value and the clump appears as an object roughly in the same direction on the plane of the sky. Thus, we combine the v$\mathrm{_{LSR}}$ derived from the HFS line profile fit of the NH$_3$(1,1) primary components and the \textit{Herschel}-based N(H$_2$) values on every position where the (1,1) line was observed with S/N\,>\,3. Then we search for clusters in this position-position-velocity-column density datacube. The importance of the secondary line components is discussed after (see Section \ref{clumppar}).

The number of clusters in the cloud was not known and we did not want to decide which clustering method to use for the proportioning beforehand. Because of this we used the \textit{NbClust} algorithm of the R statistical computing environment \citep{charrad2014}, which is able to run several types of clustering methods while varying the number of clusters to be derived. It computes so-called cluster validity indices to inspect the results of the different runs, decides what is the most appropriate number of clusters in a dataset and performs the clustering itself as well. \textit{NbClust} is able to run the k-means method (see Appendix \ref{kmeansclustering}) and hierarchical agglomerative clustering (HAC) methods with 8 different agglomeration criteria (see Appendix \ref{hac}). A more detailed description of the \textit{NbClust} package and the used indices can be found in Appendix \ref{nbclust}.

Since our parameters in the dataset are measured on different scales, standardization is necessary. We converted the position offsets to galactic offsets in arcminutes, the velocity to tenth of kms$^{-1}$ and the column density to 10$^{21}$\,cm$^{-2}$ to bring the parameter ranges closer to each other. Then we used the min-max scaling where the normalized value of parameter p$_i$ of the $n$th point in the dataset is defined by:
\begin{equation}
p_{i,n,\mathrm{norm}}=\frac{p_{i,n}-p_{i,\mathrm{min}}}{p_{i,\mathrm{max}}-p_{i,\mathrm{min}}}
\end{equation}
where p$_i$ are the 4 parameters ($\Delta$l, $\Delta$b, v$\mathrm{_{LSR}}$, N(H$_2$)) of an x=[p$_1$, p$_2$, p$_3$, p$_4$] point in the 4D parameter space with $n$ points. Min-max scaling or scaling by the range of the variables was shown to give more accurate results for the most known clustering techniques, than using traditional standardization methods, like z-score \citep{steinley2004}.

\textit{NbClust} also requires the definition of a distance metric. We used Euclidean distance metric where the distance between two points (x$_1$ and x$_2$) in the $i$=4 dimensional parameter space is:
\begin{equation}
d(x_1,x_2)=\sqrt{\sum_{i=1}^{4}(x_{1,i}-x_{2,i})^2}
\end{equation}
We can then define within-cluster and between-cluster distances for points inside one cluster and of two different clusters respectively. 

The 9 available clustering methods in \textit{NbClust} were run on the scaled and normalized TMC-1 joined dataset. When determining the ideal number of clusters with a certain clustering method, the number of clusters that was allowed to be derived varied from 2 to 15. \textbf{NbClust} computed 30 indices to measure the validity of the derived clusters (see App. \ref{nbclust}) in each case. Each index suggested the ideal number of clusters that should be derived by the method and the final decision was made by the majority rule. After this the clustering was performed. The final results of the 9 clustering methods are plotted on Fig. \ref{clumping2}.

Four clustering methods provided similar division of TMC-1 into 4 parts, see Fig \ref{clumping2}b,c,d and Fig. \ref{wardclumping}. The HAC method with the average agglomeration criterion derived 3 clusters (Fig. \ref{clumping2}e), McQuitty's criterion derived 15 parts (Fig. \ref{clumping2}a), and the HAC method with the median, centroid and single agglomeration criteria did not part the filament (Fig. \ref{clumping2}f,g,h).

In order to decide between the results of the 9 clustering methods, the sum and the weighted sum of the within-cluster variances were determined for each method. We first calculated the variance of the distances of all cluster members from the cluster centers in every cluster resulted by each clustering method with
\begin{equation}
S_{k}=\left(\langle x\rangle-x_k\right)^2 \\
\end{equation}
\begin{equation}
\sigma^2=\frac{\sum_{k}^{\,}\left(\langle S_{k}\rangle-S_{k}\right)^2}{k}
\end{equation}
where $k$ is the index of points inside a cluster, S$_{k}$ are the distances of each cluster member (x$_k$) from the cluster center ($\langle x\rangle$) in each cluster and $\sigma^2$ is the variance of these distances for each cluster. Then we calculated a sum (C$\mathrm{_m}$) and a weighted sum (C$\mathrm{_{m,w}}$), adding together the cluster variances of the clusters belonging to each clustering method with
\begin{equation}
\label{sumofvar}
C\mathrm{_m}=\sum\sigma^2
\end{equation}
\begin{equation}
\label{wsumofvar}
C\mathrm{_{m,w}}=\sum \mathrm{w}\sigma^2
\end{equation}
where w weight is the number of pixels in a cluster. The most appropriate number of clusters and the number of indices suggesting that number are found in Table \ref{nbres} for each clustering method, along with the derived sum and weighted sum of variances for the methods.

The HAC method with McQuitty's agglomeration criterion derived the clusters with the minimal sum and weighted sum of variances in the 4D-parameter space. However, when the clusters are projected back to the sky we find one too small fragment (only 3 pixels), and there are 3 fragments that are not continuous, but each of them is broken into two or three isolated parts. The existence of several very small fragments may be proved measuring several chemical species and using high spatial resolution and small spacing, but in the current study we accept the results of the k-means method as best, which had the second minimal sum of within-cluster variances.

The 4 clusters resulted by the k-means method were projected back to the plane of the sky, defining 4 objects, each of them is contiguous (see Fig. \ref{wardclumping}). Three of these sub-filaments are very elongated (\object{TMC-1F1}, \object{TMC-1F2}, \object{TMC-1F4}) and one of them (\object{TMC-1F3}) is less so. TMC-1F1 appears with the highest velocities and column densities, while TMC-1F2 on the north-eastern edge has the lowest kinetic temperature. The environment of the YSO makes up TMC-1F3. TMC-1F4 encompasses the middle and the southern regions with a range of velocities and has the highest average linewidth.

The secondary NH$_3$(1,1) line components in TMC-1F3 appear at a velocity of the neighbouring TMC-1F1. Similarly, the secondary NH$_3$(1,1) line components in TMC-1F1 appear at 5.2-5.4\,kms$^{-1}$ i.e. the velocity of TMC-1F3. The kinematics thus suggests that the clumps are slightly overlapped. Few spectra with secondary components in TMC-1F4 indicate small fragments with velocities similar to that of TMC-1F1 and TMC-1F2.

Despite the scaling and the normalization of the parameters of the joined dataset, the Euclidean distance measured in one dimension can not be unequivocally compared with the distance measured in another, since the parameters are independent and not interchangeable. For example, the min-max normalization only works well if the dispersion in each dimension is small and there are no extreme points that distort the calculations. The careful selection of the 82 spectral positions with high S/N NH$_3$ measurements and the smoothly changing N(H$_2$) ensures that this is not the case in this dataset. However, to reveal the discrepancies originating from using these statistical methods with a dataset like this, in a later study our methods should be carried out side by side with other clump-finding methods on a more extended dataset, where the decomposition of the molecular cloud is known, can be verified and the results can be compared. 

\subsubsection{The parameters and stability of the sub-filaments}
\label{clumppar}

Averaged \textit{Herschel}-based T$\mathrm{_{dust,ave}}$ were calculated inside the sub-filament contours and were found to be the same within 0.5\,K for each sub-filament. T$\mathrm{_{kin,ave}}$ kinetic temperatures of the sub-filaments were calculated as an average of the corresponding T$\mathrm{_{kin}}$ values on the positions of the 17 high S/N NH$_3$(2,2) spectra (see Table \ref{amm}) inside the given sub-filament. Apparently, T$\mathrm{_{kin}}$ shows a higher variation than T$\mathrm{_{dust}}$, that is also seen in the averaged values. Background corrected column densities were calculated for each sub-filament subtracting an average background N(H$_2$) of HCL 2. Then the total hydrogen mass was calculated integrating inside the sub-filament boundary. In our simplified assumptions the sub-filaments are considered as cylinders with L length and r radius and they are roughly perpendicular to the line of sight. L is their maximum diameter along the galactic plane and r is half of their diameter perpendicular to this. The linear mass of the sub-filaments was calculated dividing their mass with the L length. TMC-1F1 has the highest N(H$_2$)$\mathrm{_{ave}}$ value and TMC-F4 has the highest mass.

The greatest uncertainty of the mass calculation is the error in the N(H$_2$) distribution caused by the $\approx$1.5\,K uncertainty of the T$\mathrm{_{dust}}$ map. According to this, the calculated linear masses can have an error up to 56\%. We note here that the uncertainty of the T$\mathrm{_{dust}}$ values was a conservative estimate but still this error makes our virial stability consideration uncertain. Additionally, because of our sub-filaments overlap, N(H$_2$) derived from the dust emission originates from more than one structures. The ratio of N$\mathrm{_p}$(NH$_3$) derived from the secondary components and the total N$\mathrm{_p}$(NH$_3$) is around 10\%. If we assume that the relative abundance of NH$_3$ does not change much inside a sub-filament, this means that the ratio of N$\mathrm{_p}$(NH$_3$) from the main and secondary line components reflects the ratio of N(H$_2$) between the two overlapping cloud parts. Thus we can say that the contribution of the overlapping part of a sub-filament to the mass of the other one is negligible.
\begin{table}[t]
	\caption{Our sub-filaments and the coinciding objects.}
	\centering
    	\resizebox{0.5\textwidth}{!}{
		\begin{tabular}{c l}
			\hline
			Sub-filament & Other IDs  \\
\hline
TMC-1F1&\vtop{\hbox{\strut[SLF82]\,A $\mathrm{^{(1)}}$, [SLF82]\,D $\mathrm{^{(1)}}$ , [SLF82]\,F $\mathrm{^{(1)}}$,}\hbox{\strut core A $\mathrm{^{(2)}}$, TMC-1B $\mathrm{^{(2)}}$}}\\
TMC-1F2 & [SLF82]\,A $\mathrm{^{(1)}}$,[SLF82]\,D $\mathrm{^{(1)}}$,[SLF82]\,F $\mathrm{^{(1)}}$ \\
TMC-1F3 & [SLF82]\,A $\mathrm{^{(1)}}$, TMC-1X $\mathrm{^{(2)}}$\\
TMC-1F4 & \vtop{\hbox{\strut[SLF82]\,B $\mathrm{^{(1)}}$, [SLF82]\,C $\mathrm{^{(1)}}$, core C $\mathrm{^{(2)}}$,}\hbox{\strut TMC-1D $\mathrm{^{(2)}}$}} \\
\hline
		\end{tabular}}
 \tablebib{(1) \citet{snell1982}; (2) \citet{hirahara1992}.}
	\label{clustersobj}
\end{table}

We calculated the total ($\sigma\mathrm{_{total,ave}}$), the thermal ($\sigma\mathrm{_{th,ave}}$) and the non-thermal velocity dispersion ($\sigma\mathrm{_{nth,ave}}$) inside each sub-filament from the $\Delta$v HFS linewidth of their averaged spectra with equations \ref{totalsig}, \ref{thsig} and \ref{nthsig}. We note that the total velocity dispersion includes the velocity gradients inside the sub-filaments, these are 0.18, 0.15, 0.03 and 0.35\,kms$^{-1}$pc$^{-1}$ for TMC-1F1 to F4, respectively. Then the virial mass per unit length of the sub-filaments could be derived with
\begin{equation}
M\mathrm{_{lin,vir}}=\frac{2\sigma\mathrm{_{H_2,total,ave}}^2}{G} 
\end{equation}
where $\sigma\mathrm{_{H_2,total,ave}}$ is the total velocity dispersion of the hydrogen \citep{ostriker1964}. This is calculated as
\begin{equation}
\sigma\mathrm{_{H_2,total,ave}}=\sqrt{\sigma\mathrm{_{nth,ave}}^2+\sigma\mathrm{_{H_2,th,ave}}^2}
\end{equation}
 The physical parameters of the sub-filaments are listed in Table \ref{clusters}. The mass of the sub-filaments are slightly above (TMC-1F1 and TMC-1F4) or below (TMC-1F2 and TMC-1F3) their virial mass per unit length but with uncertainties around 50\% in the observed linear masses we can only safely say that all of them are close to equilibrium. The non-thermal support of the sub-filaments is generally 2-2.5 times higher than the thermal. We note that there is no significant difference between the parameters of the clusters (size, shape, masses) from the k-means and the other clustering methods that also derived 4 clusters.

\subsubsection{Sub-filaments TMC-1F1, TMC-1F2 and TMC-1F4}
\label{subf_f1_f2_f4}

Our results show that the main ridge of TMC-1 can be resolved into 3 sub-filaments. A minimal overlap of TMC-1F4 and TMC-1F2 or the presence of other small fragments is seen in the middle of the ridge in the presence of secondary line components. Otherwise these sub-filaments are quite well-separated. TMC-1F4 has the highest average turbulent velocity dispersion. The N(NH$_3$(1,1)) distribution peaks inside TMC-1F1 (see the upside down triangle in Fig. \ref{vel}b), that is also the position of the maxima of N$\mathrm{_p}$(NH$_3$) and n(H$_2$) (see Table \ref{amm}).

\citet{snell1982} observed TMC-1 in C$^{34}$S(1-0) and (2-1) and divided it into six fragments ([SLF82] A to F), many of those were overlapping. The position, extent and velocity of [SLF82] B and [SLF82] C roughly correspond to TMC-1F4. Their [SLF82]\,A, [SLF82] D and [SLF82] F coincide with the positions of TMC-1F1 and TMC-1F2, and the velocities of [SLF82] D and [SLF82] F are also similar to that of TMC-1F1 and TMC-1F2.

\citet{hirahara1992} found six cores on CCS emission maps of TMC-1. In the northern region, their core A and B (\object{TMC-1B}) coincide with the northern part of our TMC-1F1 and the overlapping part of TMC-1F3 which we detected with the secondary line components. They divide our TMC-1F4 into two parts, namely core C and D (\object{TMC-1D}) based on their CCS channel maps. We note however that they appear as one object with some structure on their CCS position-velocity and HC$_3$N integrated intesity maps. Their core E (\object{TMC-1E}) is positioned south-east of the southern edge of TMC-1F4, where we did not detect significant NH$_3$(1,1) emission. Associations of the sub-filaments and these formally known parts of TMC-1 are in Table \ref{clustersobj}.

\citet{pratap1997} mapped TMC-1 in SO(3$_2$-2$_1$), CS(2-1), and HC$_3$N(4-3) and (10-9). Their SO peak is $\approx$1$\arcmin$ north-east from our NH$_3$ peak in TMC-1F2 that we also recognize in their SO channel maps. TMC-1F4 appears as an elongated maximum of the HC$_3$N(4-3) and (10-9) integrated intensity \citep[see Fig. 7. in][]{pratap1997}. A secondary maximum is seen in HC$_3$N integrated intensity maps inside our TMC-1F1. The relative density maxima of sulfur and nitrogen-bearing molecules are in the northern sub-filaments, whereas TMC-1F4 is the relative density maximum region of the carbon-chain molecules. The cyanopolyyne peak TMC-1 CP is roughly in the middle of TMC-1F4.

TMC-1F1 and TMC-1F4 are associated with two and one 350\,$\mu $m \textit{Herschel} point sources respectively. We can not exclude that the higher T$\mathrm{_{kin}}$ measured at position \#1 (see Table \ref{amm}) is related to the nearby \textit{Herschel} point source, although we do not measure an increased temperature at position \#2. The nature of the \textit{Herschel} FIR point sources, other NIR point sources in the region \citep[see eg.][]{toth2004} and their relation to the ISM will be discussed elsewhere.

We note that the HAC method with McQuitty's agglomeration criterion defines 12 sub-filaments in the main ridge (M1 to M12 in Fig. \ref{clumping2}a). None of those have the same position, velocity and extent as any of the \citet{snell1982} or \citet{hirahara1992} fragments, but the fragment M4 which coincides with the N$\mathrm{_p}$(NH$_3$) and N(H$_2$) peak region, is located close to fragment A by \citet{hirahara1992}, although it has a greater extent.
\begin{figure}[!t]
    \includegraphics[width=\linewidth]{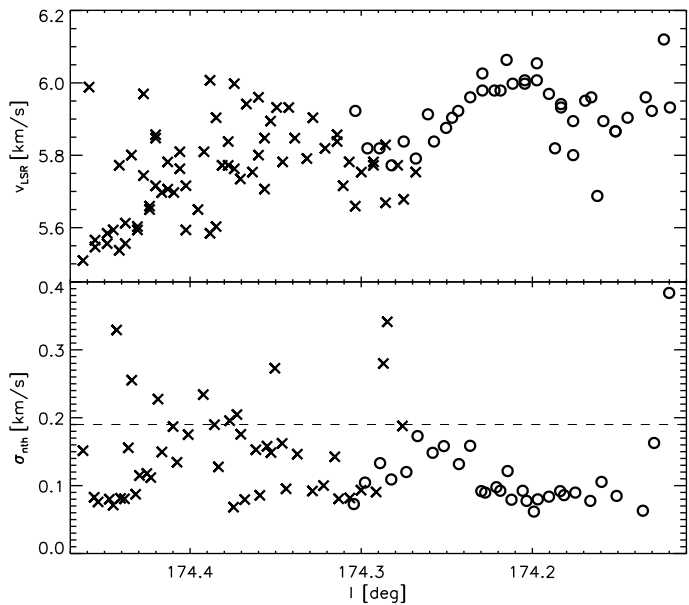}
	\caption{The central velocity and the non-thermal velocity dispersion of the NH$_3$(1,1) primary components along TMC-1F4 (crosses) and TMC-1F1 (circles) from left to right. The sound speed in the two clumps are marked with a dashed line in the lower panel.}
	\label{f1f4_vel}
\end{figure}

\subsubsection{Sub-filament TMC-1F3 and the Class\,I protostar IRAS 04381+2540}
\label{F3}

TMC-1F3 is separated from the main filament both in position and velocity. The velocity and position of a CS local maximum, \citet{snell1982} fragment A, roughly corresponds to TMC-1F3. Their Fig. 8 also indicates that TMC-1F3 partly overlaps with the main ridge. The displacement of the CS peak may be partly due to their low resolution (HPBW\,$\approx$\,2\,$\arcmin$) or a chemistry effect. TMC-1F3 corresponds to \object{TMC-1X} (identified as core X in the CCS(4-3) position-velocity maps) of \citet{hirahara1992}. It appears at similar velocities but with different extent in their HC$_3$N(5-4) integrated intensity and NH$_3$(1,1) channel maps. The overlap with the main ridge is also apparent in their figures. One may recognize TMC-1F3 in the CS and SO integrated intensity maps of \citet{pratap1997}.

TMC-1F3 was also located by \citet{yildiz2015} in $^{12}$CO with a velocity of 5.2\,kms$^{-1}$. Its position coincides with that of the protostar IRAS 04381+2540 \citep{beichman1986} that is known to drive a bipolar outflow detected in $^{12}$CO \citep{bontemps1996}, and H$_2$ \citep{gomez1997}. Its disk was resolved, and a YSO mass of 0.54\,M$_{\sun}$ was estimated by \citet{harsono2014}. \citet{apai2005} described it as a binary in formation with a jet and a substellar companion, and showed that the northern outflow cavity is opening towards us. According to \citet{yildiz2013} the velocity of the YSO is v$\mathrm{_{LSR}}$\,=\,5.2\,kms$^{-1}$ that agrees with that of our NH$_3$ clump (5.2-5.4 kms$^{-1}$). The non-thermal velocity dispersion of TMC-1F3 is the second highest and its thermal velocity dispersion is the highest of all four sub-filaments.

From the positional coincidence and velocity one may assume that the Class I protostar is related to the sub-filament. The relatively high average kinetic temperature (13\,K, see Table \ref{clusters}) of TMC-1F3 we interpret as a possible signature of the heating by IRAS 04381+2540. In fact the highest kinetic and dust temperatures were measured at the YSO position (see position \#10 with T$\mathrm{_{kin}}$\,$\approx$\,14\,K and T$\mathrm{_{dust}}$\,=\,11.4\,K in Table \ref{amm}).

The highest HFS linewidth of the NH$_3$(1,1) primary line components is at position \#8 and around it ($\Delta$v\,=\,0.39\,kms$^{-1}$, T$\mathrm{_{kin}}$\,=\,10.6\,K, see Table \ref{amm}) in TMC-1F1. It could be explained by the outflow of IRAS 04381+2540 (120$\arcsec$ south-west of \#8) releasing mechanical energy towards the main ridge. But the outflow is very compact, with a lobe extent of only around 40\,$\arcsec$ to the north and south according to \citet{hogerheijde1998} and \citet{yildiz2015}. The northern outflow lobe is blue-shifted, opening towards the observer, thus the protostar has to be positioned behind the main ridge and the outflow should extend to at least 120\,$\arcsec$ (projected distance of position \#8) in order to affect the main ridge. \citet{hogerheijde1998} calculated a dynamical age of 2100\,years from a $^{12}$CO(3-2) outflow extent of 0.02\,pc and outflow velocity of 9.3 kms$^{-1}$. One may however assume that a jet with about an order of magnitude higher speed propagated to more than 0.1\,pc in the same time interval. We note here that velocities up to 70\,kms$^{-1}$ were observed also at very low-mass YSOs \citep[see e.g. ISO 143 in][]{joergens2012}. Our assumption from the above geometry is that the protostar IRAS 04381+2540 is embedded in TMC-1F3 that may be located around 0.1\,pc behind the main ridge.

\subsubsection{Fray and fragment?}

The primary NH$_3$(1,1) line component v$\mathrm{_{LSR}}$ velocity distribution in TMC-1F1 and TMC-1F4 is continuous, as seen in Fig. \ref{f1f4_vel}. A velocity gradient of $\approx$\,0.45\,kms$^{-1}$pc$^{-1}$ is seen from SE to NW. An oscillating behaviour can be observed above the gradient in both sub-filaments, but with a higher dispersion in TMC-1F4, and the wavelength of the periodical change in TMC-1F1 from a sinusoidal approximation is $\approx$\,0.3\,pc. Similar velocity oscillations were observed in several filaments in L1517 \citep{hacar2011} together with sinusoidal column density oscillations as an evidence of core formation. The non-thermal velocity dispersions also show oscillations and are mostly sub-sonic, since the typical sound speed, c$\mathrm{_s}$=$\sqrt{\frac{k\mathrm{_B}T\mathrm{_{kin}}}{\mu m\mathrm{_H}}}$\,=\,0.19\,kms$^{-1}$ in the two sub-filaments. Supersonic points could be explained by confusion of multiple velocity components or outflows from embedded sources inside the sub-filaments. The pattern of velocity dispersions may indicate that the sub-filaments are decoupled from the large-scale turbulent velocity field and the turbulent dissipation in the cloud occurred at the scale of the sub-filaments \citep{tafalla2015}.

Comparing the position-velocity diagram of TMC-1F1 and TMC-1F4, the latter looks less regular. Analysing a dust continuum map of TMC-1, \citet{suutarinen2011} suggested that the southern half of the ridge has an extensive envelope, in contrast to the northern, i.e. the NH$_3$ peak region. They explain it as a difference in dynamical age: the northern part is more compressed. Based on time dependent modeling of the observed variation of chemistry they also concluded that the south-eastern end of the cloud is also chemically less evolved than the north-eastern end. The difference in the "chemical ages" of TMC-1F1 and TMC-1F4 was also indicated by \citet{berczik2015}.

\citet{peng1998} studied the fragmentation in a 8$\arcmin\times$8$\arcmin$ region around TMC-1 CP in TMC-1F4 with CCS measurements. They found 45 very small fragments with sizes of 0.02-0.04\,pc and masses of 0.04-0.6\,M$_{\sun}$, five of those fragments were gravitationally bound. The northern end of TMC-1F4 was resolved into cores with sizes of 0.03-0.06\,pc and masses of 0.03-2\,M$_{\sun}$ by \citet{langer1995}, based on high resolution single-dish and interferometric observations of CCS and CS.

Hierarchical fragmentation was found by \citet{hacar2013} and \citet{tafalla2015} in the nearby \object{L1495/B2013} complex. The "fray and fragment" filament evolution scenario proposed by them is the following: the filamentary cloud first fragments into intertwined, velocity-coherent filaments, so-called fibers. If the fiber can accumulate enough mass to become gravitationally unstable, it can further fragment into chains of closely-spaced cores in an almost quasi-static way. 
As seen in Table \ref{clusters}, the length of TMC-1F1, TMC-1F2 and TMC-1F4 is similar that of the fibers in L1495/B2013. The typical length of the fibers of \citet{hacar2013} is 0.5\,pc (ranging from 0.2\,pc to 1.2\,pc). TMC-1F2 even has a similar linear mass as the fiber [HTK2013]\,32 of \citet{hacar2013}. However their typical 0.5\,pc-long fibers have a factor of two lower mass then the sub-filaments of TMC-1. Some of their fibers are unstable and fragmented into chains of cores.

The velocity and linewidth distributions in TMC-1F1 in the northern part of TMC-1 is very similar to that of the fibers described by \citet{tafalla2015}. The southern part of TMC-1 with TMC-1F4 is said to be structurally and chemically less evolved than the northern part of TMC-1, also the velocity dispersion is much higher in TMC-1F4 than in TMC-1F1. But there are already small, gravitationally bound cores inside TMC-1F4. It is not clear whether "fray and fragment" is the process shaping TMC-1, because we cannot simply say that the filament first collapsed into velocity-coherent "fibers" then formed cores inside them. One has to see the distribution of small-scale structures all along TMC-1, and discuss again the "fray and fragment" scheme.

\section{Conclusions}

Our work contributes to the study of HCL 2, focusing on the TMC-1 region based on \textit{Herschel SPIRE} FIR and high S/N NH$_3$ 1.3\,cm line mapping. We found TMC-1 as one of the coldest and densest parts of HCL 2 (along with TMC-1C and HCL 2B) from our N(H$_2$) and T$\mathrm{_{dust}}$ calculations.

Our Herschel-based N(H$_2$) shows values above 10$^{22}$\,cm$^{-2}$ with two local maxima along a narrow ridge. The northern maximum is the peak N(H$_2$) position with 3.3\,$\times$\,10$^{22}$\,cm$^{-2}$. There is a third, more separated peak south-west from this. T$\mathrm{_{dust}}$ is almost constant along the higher density inner part of the ridge.

Fitting multiple HFS line profiles to the NH$_3$(1,1) spectra we identified multiple velocity components in the northern region and calculated low turbulent velocity dispersions (0.1-0.2\,kms$^{-1}$). The derived N(NH$_3$(1,1)) distribution follows the \textit{Herschel}-based N(H$_2$) distribution well. We define a new NH$_3$-peak in an offset of $\approx$\,1$\arcmin$ relative from TMC-1(NH3). The NH$_3$-based n(H$_2$) are lower than former C$^{34}$S and HC$_3$N-based values with a factor of 5-6.

Based on the observed velocity and column density variations we partitioned TMC-1 using the \textit{NbClust} package of R. The 4 derived sub-filaments have masses of 20-40\,M$_{\sun}$, three of them are quite elongated (TMC-1F1, TMC-1F2, TMC-1F4) and all of them close to gravitational equilibrium. TMC-1F1 has the NH$_3$ peak and TMC-1F4 the cyanopolyyne peak inside. TMC-1F3 partly overlaps with TMC-1F1, the Class I protostar IRAS 04381+2540 is apparently embedded into this sub-filament. The protostellar outflow is probably affecting TMC-1F1. The "fray and fragment" process is one possible scenario for the structuring of TMC-1 but the small-scale structure of TMC-1F1 should be investigated to ascertain this scenario.

A more detailed description and analysis could be given by a high resolution survey of the whole TMC-1 in carbonous and nitrogen bearing molecules (e.g. CCS and NH$_3$ lines) and then using chemical and radiative transfer models in the interpretation.

\begin{acknowledgements}
This research was partly supported by the OTKA grants K101393 and NN-111016 and it was supported by the Momentum grant of the MTA CSFK Lend\"ulet Disk Research Group. The research leading to these results has received funding from the European Commission Seventh Framework Programme (FP/2007-2013) under grant agreement No 283393 (RadioNet3). V-MP acknowledges the support of Academy of Finland grant 250741. Valuable discussions with Jorma Harju, Malcolm Walmsley and Dimitris Stamatellos are acknowledged. \\
We thank the anonymous referee for the careful reading of our manuscript and the valuable comments and suggestions. \\
Herschel was an ESA space observatory with science instruments provided by European-led Principal Investigator consortia and with participation from NASA. SPIRE was developed by a consortium of institutes led by Cardiff Univ. (UK) and including Univ. Lethbridge (Canada); NAOC (China); CEA, LAM (France); IFSI, Univ. Padua (Italy); IAC (Spain); Stockholm Observatory (Sweden); Imperial College London, RAL, UCL-MSSL, UKATC, Univ. Sussex (UK); Caltech, JPL, NHSC, Univ. Colorado (USA). This development was supported by national funding agencies: CSA (Canada); NAOC (China); CEA, CNES, CNRS (France); ASI (Italy); MCINN (Spain); SNSB (Sweden); STFC (UK); and NASA (USA). PACS was developed by a consortium of institutes led by MPE (Germany) and including UVIE (Austria); KUL, CSL, IMEC (Belgium); CEA, OAMP (France); MPIA (Germany); IFSI, OAP/AOT, OAA/CAISMI, LENS, SISSA (Italy); IAC (Spain). This development has been supported by the funding agencies BMVIT (Austria), ESA-PRODEX (Belgium), CEA/CNES (France), DLR (Germany), ASI (Italy), and CICT/MCT (Spain).\\
Based on observations with the 100-m telescope of the MPIfR (Max-Planck-Institut für Radioastronomie) at Effelsberg, the authors thank their high level technical support and assistance.
\end{acknowledgements}

\bibliography{tmc_paper}
\bibliographystyle{aa}

\begin{appendix}

\section{Column density calculation from \textit{Herschel SPIRE} measurements}
\label{herschelcol}
\subsection{Determination of the background level}
\label{bground}

The background/foreground subtraction used on the \textit{Herschel SPIRE} images after calibration with HIPE is described in Section \ref{herschelcalc}. The difference between T$\mathrm{_{dust}}$ without and with this correction is less than 0.5\,K which means 1-3\% difference in N(H$_2$) in the dense regions of TMC-1, increasing to 20-30\% in the outer regions. The contour value used to determine the background/foreground level is very similar to the baseline fitted by \citet{malinen2012} when deriving radial density profile for TMC-1 and it was mentioned by them that a background subtraction may be needed when determining filament properties.

Another way of determining the background level is calculating N(H$_2$) distribution without any correction, then taking the contour above which we were able to detect NH$_3$. This contour is around 0.7\,$\times$\,10$^{22}$ cm$^{-2}$ in our case and it marks the boundaries of the dense ISM. We created T$\mathrm{_{dust}}$ and N(H$_2$) with subtracting the average intensities found at this contour from the \textit{SPIRE} images. The difference between the T$\mathrm{_{dust}}$ values calculated with the method in the paper and this method is around 1\,K. It means a 10-20\% difference in N(H$_2$) at the dense part of the ridge, quickly increasing to more than half a magnitude difference close to the marked boundary.

\subsection{The effect of the spectral index}
\label{beta}

Early results from \textit{Planck} showed that the dust opacity spectral index, $\beta$ is around 1.8 on the all-sky map observed by \textit{Planck HFI} \citep{planck2011d}. Later the statistical analysis of the properties of the \textit{Cold Clump Catalogue of Planck Objects} using \textit{Planck} and \textit{IRAS} observations determined that in the clumps $\beta$ varies from 1.4 to 2.8 with a median value around 2.1 and it anti-correlates with temperature \citep{planck2011c}. \citet{planck2011e} examined the Taurus region and found $\beta$\,>\,1.8 in structures with temperatures below 14\,K. \citet{juvela2012b} used an average value of 2 when determining the T$\mathrm{_{dust}}$ and N(H$_2$) on the fields observed during the \textit{Herschel} Open Time Key Programme \textit{Galactic Cold Cores}. Recently \citet{juvela2015b} examined the sub-millimeter dust opacity and gave the average value of $\beta$ close to 1.9 and they indicated that it can be even higher in the coldest regions. The variations of $\beta$ were described in detail by \citet{juvela2015a}. They derived $\beta$ using different combinations of observations and concluded that it anti-correlates with T$\mathrm{_{dust}}$, correlates with N(H$_2$) and galactic latitude and its value depends on the wavelength of the observations used in the analyses. Using \textit{Herschel} measurements $\beta$\,=\,2 is appropriate for dense clumps and it can reach 2.2 in the coldest regions.

We used $\beta$\,=\,2 in the calculations in this paper but estimated the effect of changing the spectral index between 1.8 and 2.2. The differences in T$\mathrm{_{dust}}$ originating from this do not reach 10\% which gives a <\,30\% deviation in the derived N(H$_2$) distribution.

\section{Estimating N$\mathrm{_p}$(NH$_3$) from the secondary NH$_3$(1,1) velocity components}
\label{ammdens}

The physical parameters of the gas were calculated using the primary component of the NH$_3$(1,1) line, however, the secondary components also contribute to the total N$\mathrm{_p}$(NH$_3$) and N(NH$_3$(1,1)). This was estimated by first calculating the fraction of N(NH$_3$(2,2)) compared to the total N$\mathrm{_p}$(NH$_3$) from the primary velocity components with equations \ref{n11} and \ref{ntotal}. This gives typically 3-6\% for the fraction of N(NH$_3$(2,2)) in the total N$\mathrm{_p}$(NH$_3$). We then derived N(NH$_3$(1,1)) from the secondary components with equation \ref{n11}. Since the HFS of the secondary components could not be fitted well (the relative error of $\tau$ was 50\% even in the best HFS line profile fit), we assumed an optical depth of $\tau$\,=\,0.1 as upper limit which was derived from the HFS line profile fit of the noise-weighted average of all the secondary line spectra in the northern region of TMC-1. Then assuming that the contribution of the molecules in the NH$_3$(2,2) state is 10\% of this (as a conservative estimate), we computed the total N$\mathrm{_p}$(NH$_3$) from the secondary velocity components (see Table \ref{amm} in column 10).

\section{Clustering methods}
\label{appmethods}

Clustering methods are categorized based on cluster models i.e. how one defines what a cluster is. Typical models are e.g. connectivity models, where clusters are based on distance connectivity (hierarchical clustering), centroid models, where each cluster is represented by a single mean vector (k-means clustering), distribution, density or graph-based models. For each of the types there are several sub-types and different algorithms to find clusters in a dataset \citep{jain1999, halkidi2000}.

\subsection{Clustering of molecular clouds}
\label{clfind}

A well-known routine for identifying local maxima as peaks of clumps and following them to lower levels is \textsc{clumpfind} \citep{williams1994}. \textsc{clumpfind} works by searching through the data for distinct objects identified by closed contours from the highest contour levels to the lowest. Contours that surround only one peak are assigned to the corresponding clump, blended contours that surround more than one peak are split up using a "friends-of-friends" algorithm. The conservative variant of \textsc{clumpfind}, \textsc{CSAR} \citep[Cardiff Source-finding AlgoRithm;][]{kirk2013} can be used in a hierarchical mode. \textsc{gaussclump} \citep{stutzki1990, kramer1998} iteratively fits 3D Gaussian functions to a data cube to derive sources. The dendrogram technique \citep{rosolowsky2008} combines the statistical approach with segmenting the data into physically relevant structures while preserving and characterizing the hierarchy of emission isosurfaces.

\begin{figure*}[ht]
\includegraphics[width=\linewidth, trim=0 15cm 0 0]{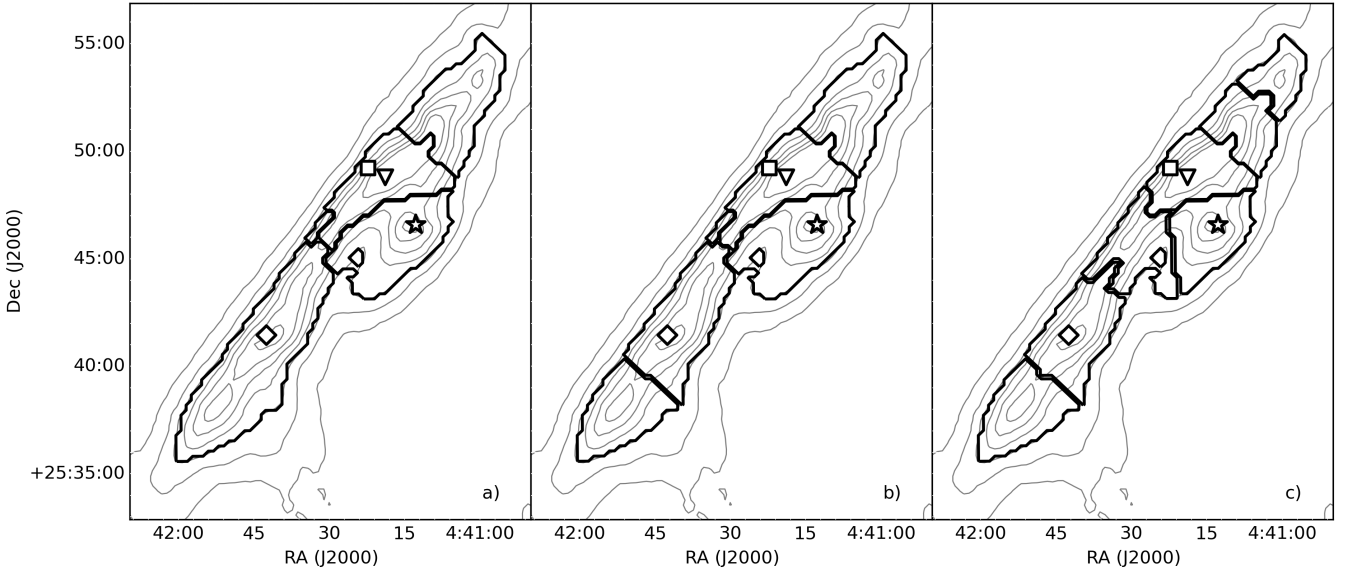}
	\caption{The results of the \textsc{clumpfind} decomposition. The algorithm was run with a: 1$\sigma$; b: 0.5$\sigma$ and c: 0.2$\sigma$ contouring stepsizes. N(H$_2$) from \textit{Herschel} and the positions of TMC-1 CP, IRAS 04381+2540, the SO peak and our NH$_3$ maximum are marked as in Fig. \ref{vel}a.}
	\label{clclumping}
    \end{figure*}

We included a decomposition of TMC-1 with \textsc{clumpfind} in Fig. \ref{clclumping} to show that the resulting clumps strongly depend on the contouring threshold and stepsize used as an input. We applied \textsc{clumpfind} with a contouring threshold of 3\,$\sigma\mathrm{_N}$ and stepsizes of 1, 0.5 and 0.2\,$\sigma\mathrm{_N}$ (where $\sigma\mathrm{_N}$ is the standard deviation measured on the N(H$_2$) map) and depending on the stepsize, resolved the filament into 4, 5 and 7 clumps. This analysis shows that decreasing the stepsize we can effectively derive clumps around every local N(H$_2$) maxima and so clump finding with \textsc{clumpfind} and similar methods on the N(H$_2$) map is not sufficient enough in the case of TMC-1.

\subsection{K-means clustering}
\label{kmeansclustering}

The k-means clustering \citep{macqueen1967, murtagh1987} uses the centroid cluster model. It is an iterative method which minimizes the within-cluster sum of squares for a given number of clusters, starting with an initial guess for cluster centers. Each observation is placed in the closest cluster, the cluster centers are updated, then the process repeats itself until the cluster centers no longer change.

\subsection{Hierarchical clustering}
\label{hac}

The hierarchical agglomerative clustering (HAC) uses distance connectivity and seeks to build a hierarchy, where each observation starts in its own cluster, then pairs of clusters are merged according to an agglomeration criterion as one moves up the hierarchy. Another approach is hierarchical divisive clustering where all observations start in one cluster, and splits are performed recursively as one moves down the hierarchy.

HAC methods require an agglomeration (or linkage) criterion. When building the hierarchy of clusters found in a dataset, the two clusters with the shortest distance between each other are merged at each step. The linkage criterion defines the way this between-cluster distance is measured. It can be defined e.g. as the distance of the elements that are the farthest away in two clusters \citep[complete-linkage;][]{sorensen1948}, the shortest distance between two elements not yet belonging to the same cluster \citep[single;][]{sokal1958}, the average distance of the elements inside two clusters \citep[average;][]{sokal1958} and the squared Euclidean distance between the gravity centers of two clusters \citep[centroid;][]{sokal1958}. Linkage criteria also can be determined with the weighted mean of the between-cluster distances \citep[McQuitty;][]{mcquitty1966}, the median of the between-cluster distances \citep[median;][]{gower1967} or with minimizing within-cluster variance \citep[Ward and its second variant;][]{ward1963, murtagh2014}.

\subsection{The \textit{NbClust} package}
\label{nbclust}

The R project for statistical computing provides a wide variety of the previously listed algorithms. We used the \textit{NbClust} package \citep{charrad2014} to select the best clustering method with the most applicable input parameters to partition the TMC-1 joined \textit{Herschel}/NH$_3$ dataset into clusters. The k-means and HAC methods with 8 different agglomeration criteria are both available in \textit{NbClust}. The package has been developed to not only perform several types of clustering methods, but to evaluate the quality of clusters (the degree with which a clustering scheme fits a specific dataset) and the optimal number of clusters (which is a required input parameter for the algorithms) in a partitioning. This is the so-called cluster validity. For this purpose, a variety of validity measures (indices) were defined in the literature. The main benefit of using this package lies in the possibility of simultaneously evaluating several clustering schemes while varying the number of clusters. This helps determining the most appropriate number of clusters for the dataset, then the resulting ideal clustering scheme can be run by the same package.

One first has to define the distance measure to be used by the clustering methods. The distance measure defines how the distance of two vectors (elements or points in the dataset) are calculated during the clustering process. Euclidean distance is the usual square distance between two vectors, Manhattan distance measures the absolute distance and maximum distance is the greatest distance of two vectors. The Canberra, binary and Minkowski distance measures are also available in \textit{NbClust} \citep{seber2009}.

\textit{NbClust} can be run on a range of cluster numbers, choosing from the measures and methods described above, while cluster validity is assessed using the validity indices. Cluster validity (evaluating the quality of clusters, optimizing the number of clusters derived) can be described using three different approaches \citep{theodoridis2006}: external criteria (comparing the results to externally provided classes or labels), internal criteria (the information obtained within the clustering process is used to evaluate how well the results fit the data) and relative criteria (compares the results with other clustering schemes resulting by the same algorithm but with different parameters). Most of the validity measures determined for the third approach \citep{milligan1985, dunn1974, rousseeuw1987} were incorporated in the \textit{NbClust} package. The validity indices combine information about within-cluster compactness, between-cluster isolation and other geometric or statistical properties of the data, the number of elements and distance measures. See \citet{charrad2014} and references therein for the detailed description of \textit{NbClust} and the definitions of distance measures, linkage criteria and cluster validity indices.

\begin{figure*}[ht]
\includegraphics[width=\linewidth, trim=0 9cm 0 0]{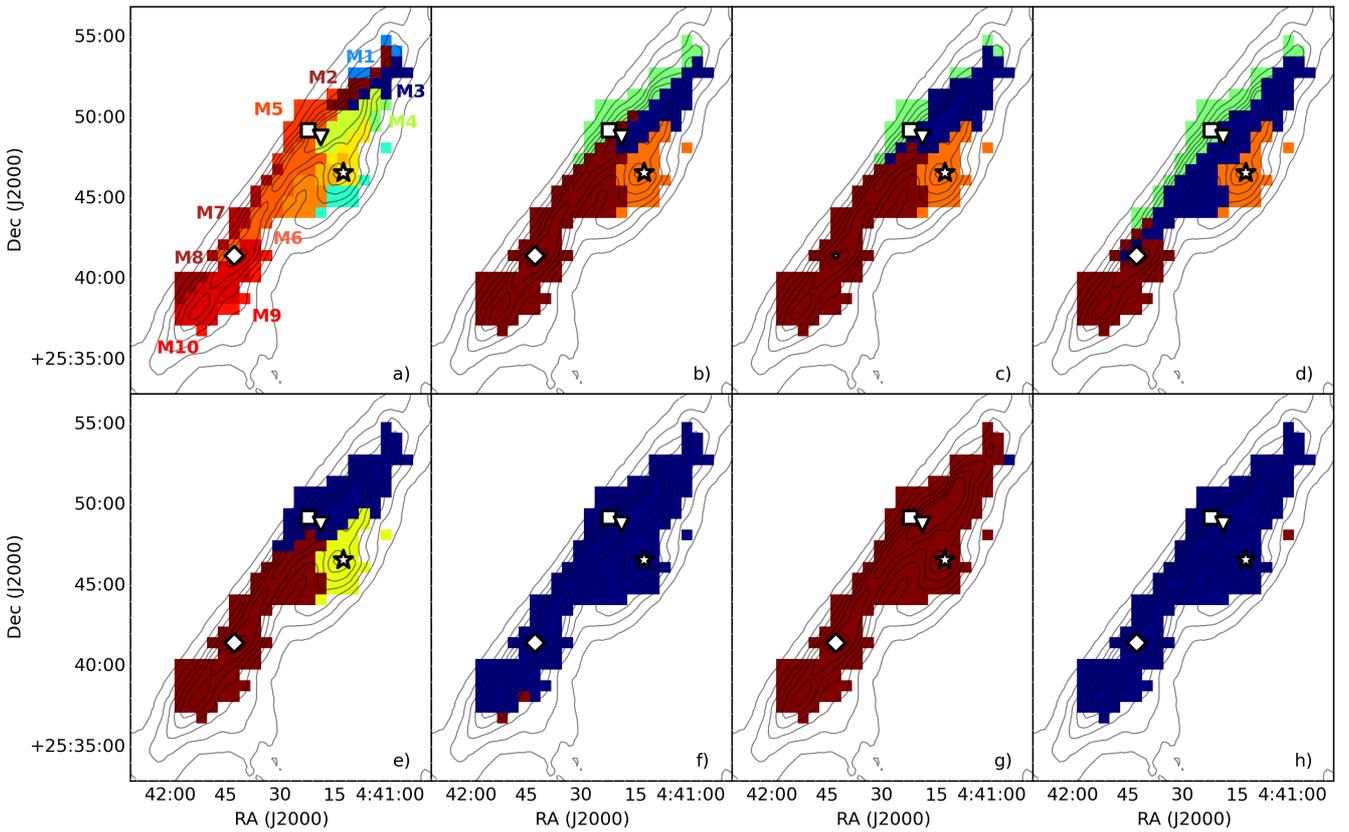}
	\caption{The best results of the other \textit{NbClust} runs: HAC methods with a: McQuitty's; b: Ward's; c: Ward's second; d: the complete; e: the average; f: the median; g: the centroid and h: the single criteria. N(H$_2$) from \textit{Herschel} and the positions of TMC-1 CP, IRAS 04381+2540, the SO peak and our NH$_3$ maximum are marked as in Fig. \ref{vel}a. The clusters in the main ridge on panel a are numbered from M1 to M10 and discussed shortly in Section \ref{subf_f1_f2_f4}.}
	\label{clumping2}
\end{figure*}
    
\end{appendix}

\end{document}